\documentclass[%
 reprint,
 superscriptaddress,
 amsmath,amssymb,
 aps,
 pra,
 floatfix,
 twocolumn
]{revtex4-2}

\newcommand{\bra}[1]{\langle #1\rvert}
\newcommand{\ket}[1]{\lvert #1\rangle}

\newcommand{\exv}[1]{\langle #1\rangle}

\DeclareMathOperator{\Tr}{Tr}

\usepackage{graphicx}% Include figure files
\usepackage{dcolumn}% Align table columns on decimal point
\usepackage{bm}% bold math
\usepackage{hyperref}% add hypertext capabilities
\usepackage{tikz}
\usepackage{dsfont}
\usetikzlibrary{calc}

\usepackage{qcircuit}

\usepackage{orcidlink}

\usepackage{amsmath}
\usepackage{upgreek}
\usepackage{physics}
\usepackage{isomath}
\usepackage{placeins}

\begin{document}

%\preprint{APS/123-QED}

% What are you doing in my latex file? Did you download this off arxiv? Think you're funnny, punk?
% Just kidding. Hopefully I didn't leave anything embarassing in this file though.

\title{Entangled Matter-waves for Quantum Enhanced Sensing}

\author{John Drew Wilson\orcidlink{0000-0001-6334-2460}}
\thanks{Corresponding author: John.Wilson-6@colorado.edu}
\affiliation{JILA, NIST, and Department of Physics, University of Colorado, 440 UCB, Boulder, CO 80309, USA}
\author{Jarrod T. Reilly\orcidlink{0000-0001-5410-089X}}
\affiliation{JILA, NIST, and Department of Physics, University of Colorado, 440 UCB, Boulder, CO 80309, USA}
\author{Haoqing Zhang\orcidlink{0000-0002-2481-2412}}
\affiliation{JILA, NIST, and Department of Physics, University of Colorado, 440 UCB, Boulder, CO 80309, USA}
\affiliation{Center for Theory of Quantum Matter, University of Colorado, Boulder, CO 80309, USA}
\author{Chengyi Luo\orcidlink{0000-0002-9470-5815}}
\affiliation{JILA, NIST, and Department of Physics, University of Colorado, 440 UCB, Boulder, CO 80309, USA}
\author{Anjun Chu\orcidlink{0000-0001-8489-9831}}
\affiliation{JILA, NIST, and Department of Physics, University of Colorado, 440 UCB, Boulder, CO 80309, USA}
\affiliation{Center for Theory of Quantum Matter, University of Colorado, Boulder, CO 80309, USA}
\author{James K. Thompson\orcidlink{0000-0003-1160-5112}}
\affiliation{JILA, NIST, and Department of Physics, University of Colorado, 440 UCB, Boulder, CO 80309, USA}
\author{Ana Maria Rey\orcidlink{0000-0001-7176-9413}}
\affiliation{JILA, NIST, and Department of Physics, University of Colorado, 440 UCB, Boulder, CO 80309, USA}
\affiliation{Center for Theory of Quantum Matter, University of Colorado, Boulder, CO 80309, USA}
\author{Murray J. Holland\orcidlink{0000-0002-3778-1352}}
\affiliation{JILA, NIST, and Department of Physics, University of Colorado, 440 UCB, Boulder, CO 80309, USA}
%
%Lines break automatically or can be forced with \\
%\collaboration{MUSO Collaboration}%\noaffiliation

\date{\today}% It is always \today, today,
             %  but any date may be explicitly specified
% PRA, PRA Letters, New Journal of Physics

\begin{abstract}
The ability to create and harness entanglement is crucial to the fields of quantum sensing and simulation, and ultracold atom-cavity systems offer pristine platforms for this undertaking.
Here, we present a method for creating and controlling entanglement between solely the motional states of atoms in a cavity without the need for electronic interactions.
We show this interaction arises from a general atom-cavity model, and discuss the role of the cavity frequency shift in response to atomic motion.
This cavity response leads to many different squeezing interactions between the atomic momentum states.
Furthermore, we show that when the atoms form a density grating, the collective motion leads to one-axis twisting, a many-body energy gap, and metrologically useful entanglement even in the presence of noise.
Noteably, an experiment has recently demonstrated this regime leads to an effective momentum-exchange interaction between atoms in a common cavity mode~\cite{MomentumExchange_Luo}.
This system offers a highly tunable, many-body quantum sensor and simulator.
\end{abstract}

{
\let\clearpage\relax
\maketitle
} 

\begin{figure*}
\centerline{\includegraphics[width=\textwidth]{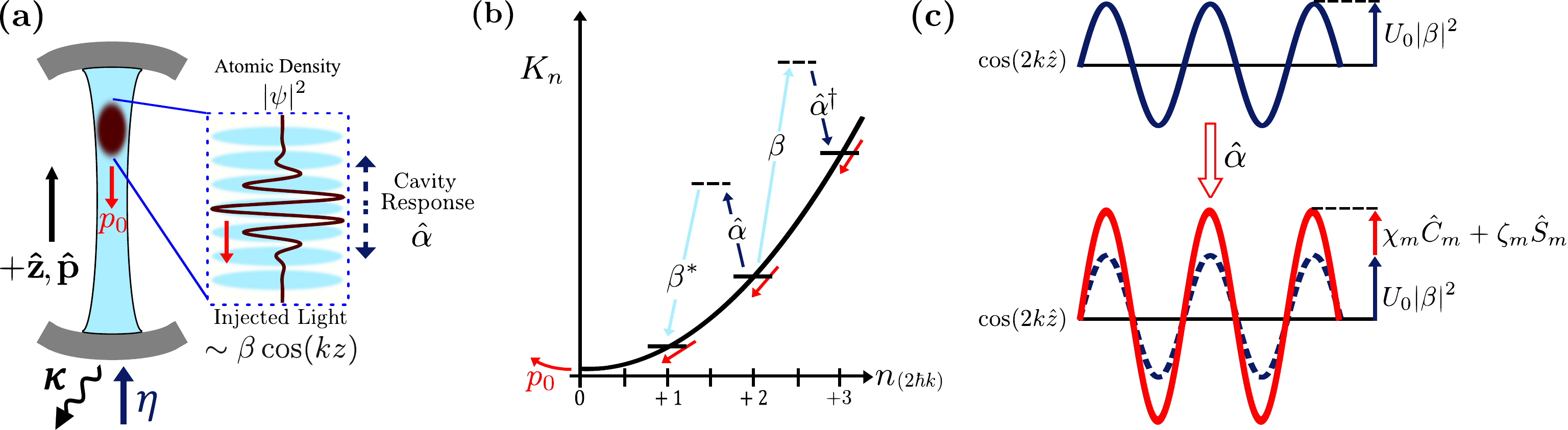}}
    \caption{
    Experimental scheme for direct generation of entanglement in momentum degrees of freedom for a matter-wave interferometer.
    (a) The atoms are well confined radially and allowed to move along $z$ with a mean momentum of $p_0$. 
    The cavity is pumped by a coherent drive such that photons enter the cavity at a rate $\eta$ and leak out at rate $\kappa$.
    As the atoms travel they interact with the injected field $\beta$ and cause a frequency shift, which in turn leads to the cavity response $\hat{\alpha}$, given in Eq.~\ref{eq:alpha}.
    (b) The quadratic kinetic energy spectrum for momentum labeled by $n$, for $p = n 2 \hbar k$, and two example interactions with cavity photons. 
    Two example processes are shown. 
    Each transition labeled by the corresponding coherent field (or photon) response for absorption of a photon, $\beta$ ($\hat{\alpha}$), and for emission, $\beta^*$ ($\hat{\alpha}^\dagger$).
    These processes act to drive the cavity field response.
    (c) Where the atoms previously felt a standing field of strength $U_0 |\beta|^2$, they now feel a standing field dependent on all other atomic momentum of strength $(U_0 |\beta|^2 + \sum_{m} \chi_m \hat{C}_m + \zeta_m \hat{S}_m )$. 
    The definition of $\hat{C}_m$ and $\hat{S}_m$ is given in text, above Eq.~\ref{eq:E_ElimModes}. The action of these operators are equivalent to the effects of a cosine and sine operator, respectively. 
    }
    \label{fig:Schematic}
\end{figure*}

For the past century, our ever-growing understanding and control of quantum mechanics has changed the way we design crucial technologies such as atomic clocks~\cite{Clock_Ludlow,Clock_Lukin}, biosensors~\cite{Bio_Taylor}, and inertial sensors \cite{Gavimetry_Qvarfort,Optomechanical_Richardson,Accelerometer_Templier}. 
These technologies have, in turn, set the stage for major advances in science, allowing for quantum mechanical tests of general relativity~\cite{RedShift_Bothwell} and direct observation of gravitational waves~\cite{LIGO_Aasi,LIGO_Tse,BlackHoles_Abbott,AdvancedBlackHoles_Abbott}.
In recent decades, it has been realized that the laws of quantum mechanics fundamentally constrain the levels of precision achievable in experiments on all scales~\cite{OpticalCoherenceText_mandel}, but also that these same laws may be used to benefit sensing experiments~\cite{HL_Holland,Squeezing_Ma,Squeeze_Ueda,LIGO_Aasi,EntangledClock_Shu}.
In particular, experiments have demonstrated growing mastery over entanglement as a resource for precision measurement platforms~\cite{SpinSqueezeQND_Kuzmich,LIGO_Tse,EntangledMomentum_Greve,EntangledModes_Malia} and quantum simulation~\cite{QSim_Georgescu,RabiHubbardSim_Mei,NuetralAtomSim_Steinert,IonSim_Cao}.

A foundational method to generate useful entanglement is that of spin squeezing~\cite{EntangledClock_Shu} where atom-atom correlations decrease uncertainty in one phase quadrature by increasing noise in an orthogonal quadrature~\cite{Squeeze_Ueda}. 
This allows sensing protocols to surpass the standard quantum limit (SQL) dictated by wavefunction collapse of unentangled particles, which bounds the uncertainty of an inferred parameter $\phi$ to $\Delta \phi^2 = 1/N$ for $N$ constituent particles.
Spin squeezing is primarily done using fine or hyperfine states of atoms~\cite{Leroux,PedrozoPenafiel,DFS_Reilly}, which require internal level dynamics to play a major role.
Oftentimes, this means one must transfer the entanglement to another degree of freedom prior to sensing.
This is often the case for atomic clocks~\cite{EntangledClock_Shu} where one requires an optical clock transition, and in gravity gradiometry where one requires spatial or kinetic separation.

If one could directly create entanglement among matter-waves similar to a spin-squeezing process, it would create the opportunity for measurements of gravitational effects through matter-wave interferometry with a true quantum advantage.
So far however, generating entangled matter-waves has always involved atomic internal levels as a proxy for the necessary dynamics~\cite{EntangledMomentum_Greve,EntangledModes_Malia}.
Past theoretical works have suggested a myriad of techniques to couple spin squeezing dynamics to the momentum states, such as through the recoil due to absorption and emission of photons~\cite{SU4_Wilson,OptimalGenerators_Wilson},  two counter-propagating modes of a ring cavity~\cite{Bragg_Shankar}, or by using conditional measurements on electronic states to replicate spin squeezing on momentum states in a ring cavity~\cite{MomentumSqueezing_Guglielmo}.

In this Letter, we present a method for achieving momentum based entanglement in a manner that does not rely on interactions between electronic states of an atom.
The mechanism enabling this entanglement generation is the shift in the cavity photon frequency in response to the atomic motion through the cavity.
This cavity response is reflected back on the atoms as a dipole force, thereby allowing atoms to self-interact and creating the possibility for a different paradigm of entanglement enhanced matter-waves.
We derive these interactions from a general atom-cavity model already realized in many experiments~\cite{TCQED1999_Ye,TCQubits_Fink}.
Furthermore, we show that when the atoms form a density grating, their collective motion leads to a pseudo-spin exchange interaction that may be used to realize OAT dynamics and a many-body energy gap.
In this limit, we derive the One-Axis Twisting (OAT) dynamics between momentum states only through cavity mediated exchange interactions which were recently observed in an experiment~\cite{MomentumExchange_Luo}.
We show that these exchange interactions may be used to generate metrologically useful entanglement even in the presence of noise, which could be used to surpass the SQL in matter-wave interferometry.
Notably, entanglement is generated between solely the momentum degrees of freedom, which differs from previous works~\cite{EntangledMomentum_Greve,SU4_Wilson,OptimalGenerators_Wilson,MomentumSqueezing_Guglielmo,SpinExchange_Norcia,PhaseTransition_Muniz}.

\noindent \emph{Self-Interaction via Cavity Frequency Modulation}. ---
In order to study the effects that drive entanglement, we consider an atomic cloud in an optical cavity with one relevant mode, as shown in Fig.~\ref{fig:Schematic}(a).
There are $N$ atoms of mass $m$ and two relevant excited and ground internal states, labeled $\ket{e}$ and $\ket{g}$ respectively, that are separated by an optical frequency $\omega_a$.
The atoms travel along the cavity with mean momentum $p_0$, and are spatially dilute enough that atom-atom interactions may be ignored compared to the interactions with the common cavity mode.
The cavity mode has frequency $\omega_c$, which is far detuned from the atomic transition.
The cavity mode has a wavenumber $k$ and the atoms have a recoil frequency of $\omega_r = \hbar k^2 / 2 m$ due to absorption and emission of cavity photons.
At the anti-node, the atoms have a maximum coupling of $g$ to the cavity.
The cavity is pumped by a coherent laser with amplitude $\eta$ and frequency $\omega_p$, which is also far detuned from the atomic transition.
The cavity decays at rate $\kappa$, while the atoms spontaneously emit light into free space at a rate $\gamma$.

The system's density matrix evolves under the Born-Markov master equation, $ \dot{\hat{\rho}} = - i[ \hat{H}, \hat{\rho} ]/\hbar + \sum_j \hat{\mathcal{D}} [\hat{L}_j] \hat{\rho}$, where $\hat{H}$ is the system Hamiltonian, $[\hat{A},\hat{B}] = \hat{A} \hat{B} - \hat{B} \hat{A}$ is the commutator, the Lindbladian superoperator is given by $\hat{\mathcal{D}} [ \hat{O} ] \hat{\rho} = \hat{O} \hat{\rho} \hat{O}^{\dagger} - ( \hat{O}^{\dagger} \hat{O} \hat{\rho} + \hat{\rho} \hat{O}^{\dagger} \hat{O} ) / 2$ summed over all jump operators $\hat{L}_j$.
Prior to any simplifications, the system Hamiltonian is given by the Tavis-Cummings model~\cite{TavisCummings,dong_TavisCummings} with spatial dependence, $\hat{H}_\mathrm{org} = \hat{H}_e + \hat{H}_c + \hat{H}_k$ where $\hat{H}_e = \hbar \Delta_a \sum_{\mu=1}^N \hat{\sigma}^z_\mu/2$ is the atomic energy and $\hat{H}_c = \hbar \Delta_c \hat{a}^\dagger \hat{a} + \hbar  \eta \left( \hat{a} + \hat{a}^\dagger \right)$ is the cavity energy and pump contribution. Lastly, $\hat{H}_k$ is the term containing atomic motion and the atom-cavity coupling:
\begin{equation} \label{eq:Hinit}
\hat{H}_{k} =\sum_{\mu=1}^N \frac{\left(\hat{p}_\mu - p_0 \right)^2}{2m} + \hbar g \cos(k \hat{z}_\mu) \left( \hat{\sigma}^+_\mu \hat{a} + \hat{a}^\dagger \hat{\sigma}^-_\mu  \right)
\end{equation}
where $\hat{a}^{\dagger}$ ($\hat{a}$) is the photon creation (annihilation) operator of the cavity mode, $\hat{p}_\mu$ ($\hat{z}_\mu$) is the momentum (position) operator for the $\mu$th atom along to the cavity axis, and $\hat{\sigma}^z_\mu=\ket{e}_\mu\bra{e}_\mu-\ket{g}_\mu\bra{g}_\mu$, $\hat{\sigma}^+_\mu=(\hat{\sigma}^-_\mu)^{\dagger}=\ket{e}_\mu\bra{g}_\mu$ are Pauli matrices for the $\mu^{\text{th}}$ atom.
The frequencies, $\Delta_c \equiv \omega_c - \omega_p$ and $\Delta_a \equiv \omega_a - \omega_p$ are the pump frequency detuning from the cavity photon frequency and the atomic transition frequency, respectively.
Decoherence is due to cavity decay and spontaneous emission: $\hat{L}_\mathrm{cav} =  \sqrt{\kappa} \hat{a}$, $\hat{L}_{\mathrm{sp},\mu}(\theta) = \sqrt{\gamma} \ \hat{d}_\mu(\theta) \hat{\sigma}^-_\mu$, where $\hat{d}_\mu(\theta) = \sqrt{\mathcal{N}(\theta)} \exp[-i k \hat{z}_\mu \cos(\theta)]$ represents the atomic recoil following the dipole radiation pattern $\mathcal{N}(\theta)$, for $\theta$ being the angle between the emitted photon and the positive $z$-axis~\cite{Doppler_Castin,Thesis_Bartolotta}, shown in the Supplemental Material (SM)~\cite{suppMat}.

To recover momentum-only dynamics, we assume that the atoms all start in the ground state and use the fact that the cavity photon frequency and the laser pump frequency are both far detuned from the atomic transition frequency, such that $\abs{\omega_a - \omega_c},\ \abs{\Delta_a} \gg \sqrt{N} g$.
This means that any dynamics involving the excited state are rapid compared to the ground state dynamics.
As a result, we may consider the absorption of a photon and subsequent re-emission to be nearly instantaneous on the timescales of the atomic motion, thereby allowing us to adiabatically eliminate the excited state~\cite{suppMat}.
After elimination, the atom-field Hamiltonian is
\begin{equation}
\label{eq:E_Elim}
\hat{H}_\mathrm{af} = \hat{H}_c + \sum_{\mu=1}^N \left( \frac{\left[ \hat{p}_\mu - p_0 \right]^2}{2 m} - \hbar U_0 \cos(2 k \hat{z}_\mu) \hat{a}^\dagger\hat{a} \right)
\end{equation}
where, $\Delta_c' \equiv \Delta_c - N U_0$ is now the pump detuning from the dressed cavity photon frequency, and $U_0 \equiv ( \Delta_a / 2 ) g^2 / (\Delta_a^2 + \gamma^2/4)$ is the effective coupling strength between the atomic motion and the cavity field.
Now, spontaneous emission from atom $\mu$ leads to $\hat{L}(\theta)_\mu = \sqrt{2 \gamma U_0/\Delta_a } \ \hat{a} \cos( k \hat{z}_\mu )\hat{d}_\mu(\theta)$.
The spontaneous emission is now understood as a photon initially being absorbed from the cavity field through $\hat{a} \cos(k \hat{z}_\mu)$ at rate $2 U_0/\Delta_a$, and subsequently emitted through $\hat{d}_\mu(\theta)$ at rate $\gamma$.

The relevant physical mechanism that leads to entanglement is the spatially dependent frequency shift on the cavity induced by the atomic motion.
From Eq.~\eqref{eq:E_Elim}, we can see that the rapid excited state dynamics cause a two-photon recoil, shown in Fig.\ref{fig:Schematic}(b), which now contributes a frequency shift to the cavity photons. 
This means the frequency of the cavity photons now depends on the position of the atoms, via $( \Delta_c' - U_0 \sum_{\mu=1}^N \cos(2 k \hat{z}_\mu) ) \hat{a}^\dagger \hat{a} $.
As the atoms move, the position dependent frequency shift oscillates between a red and blue shift on the cavity photons, whereupon this cavity response reflected back on the atoms as a dipole force.

This process couples discrete families of momentum states, where each state is by $2 \hbar k$. 
The two-photon recoil now only couples neighboring states in these momentum families, i.e.~$\ket{p}_\mu$ is only coupled to $\ket{p\pm 2 \hbar k}_\mu$.
As a result, we consider a generalized treatment of these momentum families through a second quantization approach~\cite{suppMat} in order to study these dynamics.
Momentum states are separated by integer multiples, $n 2 \hbar k$ for integer $n$, where the annihilation (creation) operator for an atom in the state $\ket{n 2 \hbar k+p_0}$ is $\hat{b}_n$ ($\hat{b}^\dagger_n$).
This requires that the atoms initially have a narrow spread in momentum about $p_0$~\cite{suppMat} less than one $\hbar k$.
On this discrete momentum family, the two photon recoil becomes quadratic in the mode operators between neighboring momentum states; $\hat{J}^\pm_n \equiv \hat{b}^\dagger_{n\pm1}\hat{b}_{n}$ such that $\sum_\mu \mathrm{e}^{\pm i 2 k \hat{z}_\mu} = \sum_n \hat{J}^\pm_n$.
Furthermore, because cosine is the sum of two exponential terms we may rewrite it as a sum of raising and lowering operators; $\sum_\mu \cos(2 k \hat{z}_\mu) = \sum_{n} \hat{C}_n$ where $\hat{C}_n \equiv ( \hat{J}^+_n + \hat{J}^-_{n+1} )/2$. Similarly, $\sum_\mu \sin(2 k \hat{z}_\mu) = \sum_{n} \hat{S}_n$ where $\hat{S}_n \equiv ( \hat{J}^+_n - \hat{J}^-_{n+1} )/(2i)$.
Lastly, the kinetic energy simplifies to $\sum_{\mu=1}^N \left( \hat{p}_\mu - p_0\right)^2/(2\hbar m) = \sum_{n} K_n \hat{b}^\dagger_n \hat{b}_n$, where $K_n = 4 \omega_r (n-n_0)^2$ is the kinetic energy for the $n^\mathrm{th}$ momentum state and $n_0$ is defined such that $p_0 = n_0 2 \hbar k$.
This allows us to simply express Eq.~\ref{eq:E_Elim} as
\begin{equation}\label{eq:E_ElimModes}
\hat{H}_\mathrm{af} = \hat{H}_c + \sum_{n} \hbar K_n \hat{b}_n^\dagger \hat{b}_n - \hbar U_0 \hat{C}_n \hat{a}^\dagger\hat{a}.
\end{equation}

It is clear that the cavity dynamics depend only on the pump and the atomic motion.
The pump injects a coherent field into the cavity, causing the cavity mode to become displaced by an amount $\beta \equiv \eta/(-\Delta_c' + i \kappa/2)$.
The effects of atomic motion on the cavity mode are succinctly captured via the coupling $U_0 \sum_{n} \hat{C}_n \hat{a}^\dagger \hat{a}$---which is the the cosine dependent frequency shift contributed per momentum mode $n$.
This means the cavity response can be separated into two components, a classical field injected by the pump, $\beta$, and the photons that interact with the atoms through a small coupling frequency $U_0$.
We solve the equations of motion for these photons by adiabatically eliminating~\cite{suppMat} the cavity response following Ref.~\cite{Lindblad_jager},
which yields the photon motion exclusively in terms of the operators describing the atomic motion.
Beginning to end, this process yields the effective replacement $\hat{a} \rightarrow \beta + \hat{\alpha}(t)$ where $\hat{\alpha}(t)$ replaces the annihilation operator in order to describe the photon motion, and only depends on the operators for atomic motion:
\begin{equation}\label{eq:alpha}
\hat{\alpha}(t) \equiv \beta \sum_{n} \left(  A^+_n \hat{J}^+_{n} + A^-_n \hat{J}^-_{n} \right)/2,
\end{equation}
where the amplitude $A^\pm_n \equiv U_0/(\Delta_c' + ( K_{n\pm1} - K_n ) - i \kappa/2 )$ depends on the required change in kinetic energy due to the two photon recoil.
\begin{figure}
\centerline{\includegraphics[width=\linewidth]{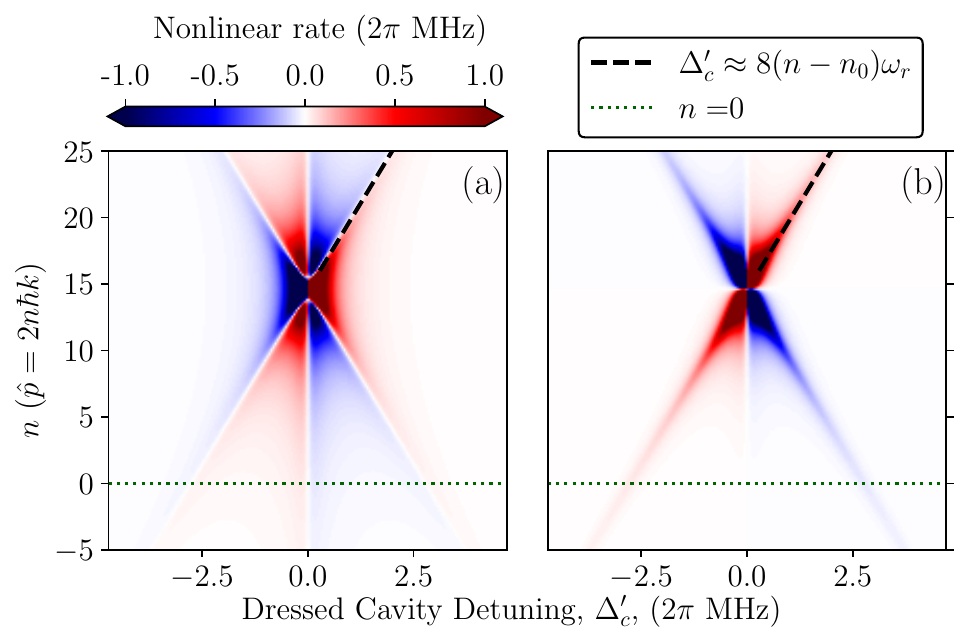}}
 \caption{
 (a) $\chi_n$ for all momentum states $n$.
 (b) $\zeta_n$ for all momentum states $n$.
 For both plots, $n_0 = 15$ sets the resonance.
 The black dashed line represents one of the resonances in $\chi_n$ for $\Delta_c' = 8(n-n_0)\omega_r$.
 There is a corresonding line of negative slope for the other resonance, not explicitly labeled.
 The green dotted line for $n=0$, denotes the rate $\chi_0$, matching the one for the momentum states prepared in Ref.~\cite{MomentumExchange_Luo}.
 The cut off of $2 \pi$ MHz was chosen to show contrast near $n=0$, but not cut off the features near $n_0$.
 }
 \label{fig:Fig2}
\end{figure}
This yields a Hamiltonian that only involves the classical injected field, and quantum momentum degrees of freedom:
\begin{equation}
\begin{aligned} \label{eq:Hatom}
\frac{\hat{H}_{a}}{\hbar} =& \sum_{n} \left( K_n \hat{b}_n^\dagger \hat{b}_n - U_0 |\beta|^2 \hat{C}_n \right) \\
& - \sum_{n,m} \left[ \chi_m \hat{C}_m + \zeta_m \hat{S}_m \right] \hat{C}_n,
\end{aligned}
\end{equation}
where $\chi_n \equiv  \frac{U_0 |\beta|^2 }{2} \mathfrak{R}( A^+_n + A^-_{n+1} )$ and $\zeta_n \equiv  \frac{U_0 |\beta|^2 }{2} \mathfrak{I}( A^-_{n+1} - A^+_n )$ are the rates for the cosine and sine non-linear interactions, with $\mathfrak{R}(Z)$ and $\mathfrak{I}(Z)$ being the real and imaginary parts of $Z$, respectively.

The jump operators are now given by the same collective atomic recoil due to cavity decay, now $\hat{L}_c = \sqrt{\kappa}\ \hat{\alpha}$, and a single atomic recoil term, which is not simply expressed in this formalism, but is treated explicitly in the SM~\cite{suppMat}.
The two decay processes are understood simply as follows: the collective atomic recoil occurs when a photon entangled to the atomic motion is lost out of the cavity, through $\hat{\alpha}$ with rate $\kappa$. The individual atomic recoil, meanwhile, occurs when an atom directly scatters a photon from the coherent field, at rate $2 U_0 |\beta|/\Delta_a$, out of the cavity at angle $\theta$, at rate $\mathcal{N}(\theta)\gamma$.

The two non-linear terms, $\chi_m \hat{C}_m \hat{C}_n$ and $\zeta_m \hat{S}_m \hat{C}_n$, generate generalized squeezing dynamics~\cite{SUN_Yukawa}.
The cosine-cosine non-linearity, $\chi_m \hat{C}_m \hat{C}_n$, is the modification of the optical lattice due to spatially dependent frequency shift of the cavity mode, i.e.~what was previously $U_0 \hat{C}_n \hat{a}^\dagger \hat{a}$ in Eq.~\ref{eq:E_ElimModes}.
The sine-cosine non-linearity, $\zeta_m \hat{S}_m \hat{C}_n$, is due to the fact that this frequency shift causes the cosine and sine quadrature to no-longer be perfectly orthogonal, and we note $\zeta_m$ is nearly zero unless near a resonance in $\chi_m$.
As a function of momentum and pump detuning from the dressed cavity photon frequency, the two rates $\chi_n$ and $\zeta_n$ are shown in Fig.~\ref{fig:Fig2} with the physical parameters given in the SM~\cite{suppMat}, where one of the two resonances is shown as a black dashed line.

In the rest of this paper we study the entanglement generation in a simple regime of Eq.~\eqref{eq:Hatom}, where we restrict to only two momentum states.
There, the non-linear interactions in Eq.~\eqref{eq:Hatom} yield OAT dynamics on the two-level manifold, which we show leads to an effective momentum exchange~\cite{MomentumExchange_Luo} and metrologically useful entanglement.

\begin{figure} \centerline{\includegraphics[width=\linewidth]{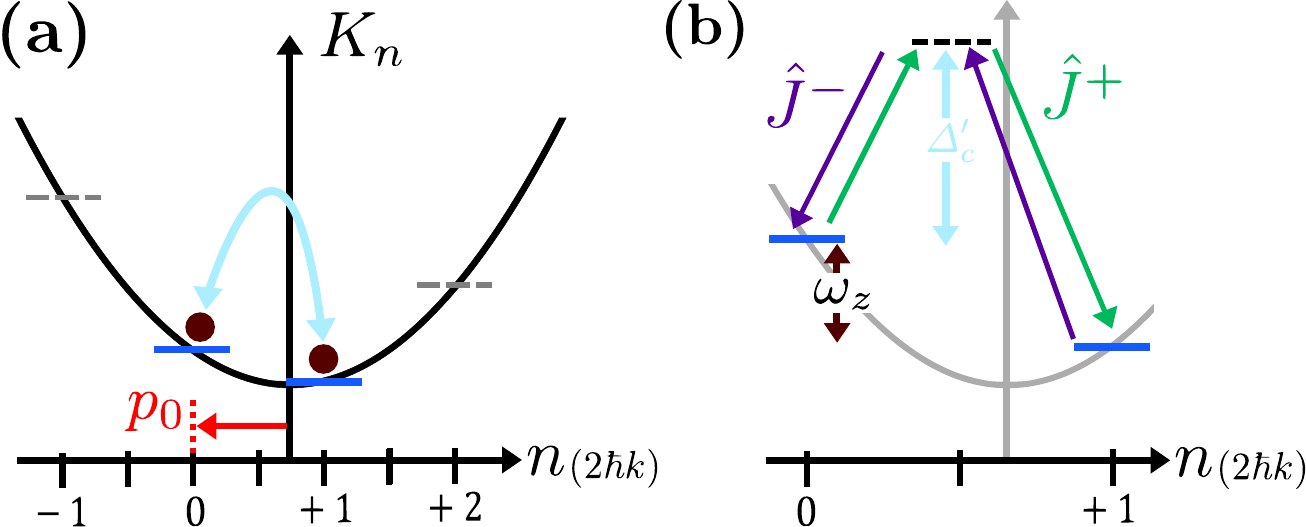}}
 \caption{
 (a) The kinetic energy spectrum, with the translation due to $p_0$ shown in red.
 In order to prepare the atoms into a collective density grating, a Bragg pulse is applied driving the transition from the $0 \hbar k - p_0$ to the the $2 \hbar k - p_0$ state prior to the cavity pump being turned on, so each atom starts in $\ket{\psi(t=0)} = (\ket{0}+\ket{1})/\sqrt{2}$.
 (b) The kinetic energy spectrum zoomed in to only show the two-level approximation.
 Here, we've shown $p_0 \approx \frac{1}{3} 2 \hbar k$.
 In the two-level approximation, the effects of the pseudo-spin raising (green) and lowering (purple) operators are shown with energy $\omega_z$ between $0$ and $+2$, and the meta-state (dotted line) with the energy of a dressed cavity photon, $\Delta_c'$.
 }
 \label{fig:Fig3}
\end{figure}

\noindent \emph{Formation of a Density Grating}. ---
When the atoms occupy just the two momentum states, $n=0$ and $n=+1$. they form a density grating.
This may be achieved physically by selecting an appropriate $p_0$ and applying a Bragg pulse.
This process is shown schematically in Fig.~\ref{fig:Fig3} (a), and symmetrically prepares each atom in the initial state $\ket{\psi(t=0)} = (\ket{0}+\ket{1})/\sqrt{2}$
The choice of $p_0$ translates the kinetic energy spectrum in momentum, shown in Fig.~\ref{fig:Fig3} (a), thereby allowing the momentum states $0 \hbar k$ and $+2 \hbar k$ to be addressed independently by tuning the dressed cavity-pump detuning, $\Delta_c'$, as is shown in Fig.~\ref{fig:Fig3} (b).

By restricting to these two-levels, we are left with only two-level operators shown schematically in Fig.~\ref{fig:Fig3} (b). 
For brevity, we drop the subscripts; $\hat{J}^+ \equiv \hat{J}_0^+$ and $\hat{J}^- \equiv \hat{J}_1^-$, and $\hat{J}_z \equiv \frac{1}{2} ( \hat{b}^\dagger_1 \hat{b}_1 - \hat{b}^\dagger_0 \hat{b}_0 )$.
These two momentum states are separated by a kinetic energy $ \hbar \omega_z \equiv \hbar ( K_{+1} - K_0 ) $.
The interference pattern of the two momentum states forms a density grating composed of all $N$ atoms.
The motion of this density grating modifies the dressed cavity frequency, and drives scattered photons according to
$ \hat{\alpha}(t) \approx \frac{\beta}{2} \left( A^+_0 \hat{J}^+ + A^-_{+1} \hat{J}^- \right)$,
with the now collective amplitudes $A^\pm = U_0/(\Delta_c' \pm \omega_z - i \kappa/2)$.
These substitutions give a typical OAT Hamiltonian;
\begin{equation}
\begin{aligned} \label{eq:su2}
\frac{\hat{H}_{\mathrm{eff}}}{\hbar} =& \omega_z \hat{J}_z - U_0 |\beta|^2 \hat{J}_x - \chi \hat{J}_x^2 - \frac{\zeta}{2} \left( \hat{J}_x \hat{J}_y + \hat{J}_y \hat{J}_x \right), \\
& \hat{L}_c = \sqrt{\kappa |\beta|^2 / 4} \ ( A^+_0 \hat{J}^+ + A^-_{+1} \hat{J}^- ),
\end{aligned}
\end{equation}
where $\hat{J}_x \equiv (\hat{J}^+ + \hat{J}^-)/2$ and $\hat{J}_y \equiv (\hat{J}^+ - \hat{J}^-)/2i$ have taken the place of cosine and sine, respectively. 
For clarity, non-linear rates on this two-level manifold simplify to
$\chi \equiv \chi_0 = U_0 |\beta|^2 \mathfrak{R}( A^+_0 + A^-_{+1} )/2$ and
$\zeta \equiv \zeta_0 = U_0 |\beta|^2 \mathfrak{I}( A^-_{+1} - A^+_0 )/2$.
Notably, $\chi$ is the same as the spin exchange rate given in Ref.~\cite{MomentumExchange_Luo}, and matches the slice at $n=0$ of Fig.~\ref{fig:Fig2} (a).
The two non-linearities in Eq.~\eqref{eq:su2} arise from the cosine-cosine and cosine-sine modulation in Eq.~\eqref{eq:Hatom}, where now the motion of the atomic ensemble acts in unison rather than in many different momentum states, leading to an effective collective enhancement of $N$ on the non-linear rates.

With these rates defined, we now discuss the conditions for energetically isolating these two momentum states from the rest of the momentum basis.
We require that higher momentum states aren't excited via either Bragg scattering off the coherent field, $\hat{C}_n$, or non-linear interactions, $(\chi_m \hat{C}_m + \zeta_m \hat{S}_m)\hat{C}_n$ in Eq.~\eqref{eq:Hatom}.
These restrictions ensure the dynamics remain in the two-level manifold, and give limits on the rates $U_0 |\beta|^2$, $\chi$ and $\zeta$, in Eq.~\eqref{eq:su2};
\begin{equation} 
\label{eq:Restrictions}
\abs{U_0} \abs{\beta}^2 \ll \frac{4 k p_0}{m}, \quad \mathrm{and} \quad N \abs{\chi},\ N \abs{\zeta} \ll 8 \omega_r.
\end{equation}
The explicit calculation of these is given in the SM~\cite{suppMat}, and it is important to note that the inequalities in Eq.~\eqref{eq:Restrictions} only guarantees unitary dynamics won't drive higher momentum states-- quantum jumps may still drive higher momentum states.
The first of these terms is the condition that the atoms are well into the conduction band of the optical lattice--meaning they translate smoothly over the lattice.
The second term is the condition that the non-linear effects aren't driven so strongly that they couple to higher momentum states.

Lastly, we consider the regime where the atoms are fast moving, or the pump is far detuned from the cavity.
In Fig.~\ref{fig:Fig2} we see that either of these conditions, large $p_0$ or large $\Delta_c'$, will yield a small $\chi$ and $\zeta$.
Here, $\omega_z$ becomes large compared to all other frequencies, such as $U_0 |\beta|^2$, $N \chi$, and $N\zeta$.
We take an interaction picture with $R = \text{exp}(i \omega_z \hat{J}_z t)$ and drop terms which are fast rotating.
Mathematically, this is a rotating wave approximation and physically, this means the rate of single atom interactions with the optical lattice is small compared to the kinetic energy.
This means the raising operators transform to $\hat{J}^\pm \mathrm{e}^{\pm i \omega_z t}$, so that $R (\hat{J}_x \hat{J}_y + \hat{J}_y \hat{J}_x) R^\dagger \approx 0$, and $4 R(\hat{J}_x)^2 R^\dagger \approx \hat{J}^+ \hat{J}^- + \hat{J}^- \hat{J}^+ = 2 (\hat{J}_z)^2 - 2 \hat{J}^2$.
Lastly, the collective decay in the Born Markov master equation splits such that $\hat{\mathcal{D}}[R\hat{L}_c R^\dagger] = \hat{\mathcal{D}}[\hat{L}_c^+] + \hat{\mathcal{D}}[\hat{L}_c^-]$, where  $\hat{L}_c^\pm = \sqrt{\Gamma_\pm}\ \hat{J}^\pm$, with $\Gamma_\pm = \kappa |\beta|^2 |A^\pm|^2 / 4$.

This regime represents an effective momentum exchange between the atoms, and this interaction drives the effects studied in Ref.~\cite{MomentumExchange_Luo}, where
the Hamiltonian simplifies to
\begin{equation}\label{eq:OAT}
\begin{aligned}
& \hat{H}_\mathrm{OAT}= \hbar \chi \left[ \hat{J}^2 - ( \hat{J}_z )^2 \right]/2.
\end{aligned}
\end{equation}
In this regime, only the energy conserving interactions of $\hat{J}^+ \hat{J}^- + \hat{J}^- \hat{J}^+$ affect the unitary dynamics, where one atom exchanges its momentum state with another via a dispersive cavity photon.
Through this energy conservation, a strong bias appears against effects which would take the atoms out of a symmetric configuration.
This leads to a large many-body energy gap of the form $\hat{J}^2$, which is a pseudo-spin length that biases against dynamics between states with different $\hat{J}^2$ eigenvalues, i.e., between states of different permutational symmetry.
This many-body energy gap leads to a further suppression of single atom effects by slowing down the effective rate of single particle decoherence effects such as Doppler broadening~\cite{MomentumExchange_Luo,GapProtection_SchleierSmith}.

\begin{figure}    
\centerline{\includegraphics[width=\linewidth]{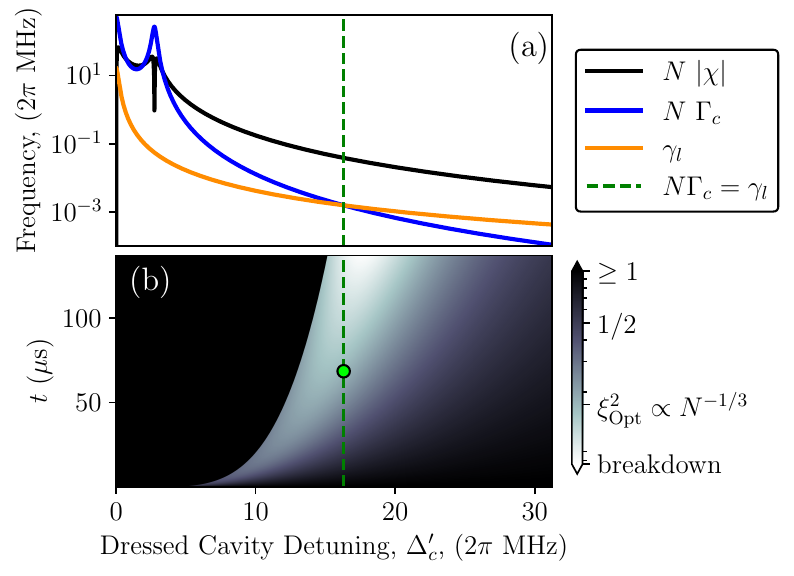}}
    \caption{
    (a) Collective rates that appear in the optimized spin squeezing parameter.
    (b) Eq.~\eqref{eq:SqueezeParam} as a function of squeezing time, $t$, and Dressed cavity detuning, $\Delta_c'$.
    We note that while values of $\xi^2$ lower than $\xi^2_\mathrm{opt}$ are found, these values are in the regime where the assumption that noise adds in quadrature is close to breaking down, i.e. where the assumptions that $N\Gamma_+ t$, $N\Gamma_+ t\ll1$ are close to breaking down.}
\label{fig:Fig4}
\end{figure}

\emph{Spin Squeezing with Decoherence}. ---
To calculate the usefulness in the presence of noise we consider the atomic species \textsuperscript{87}Rb,
We use the numerical parameters given in the SM~\cite{suppMat}.
Collective emission through the cavity is the biggest barrier to achieving a quantum advantage through this momentum squeezing, unless the pump is far detuned cavity, whereupon single particle emission becomes relevant. 
This causes atoms to be lost from the two-level manifold at a rate $\gamma_l \equiv (\gamma U_0 |\beta|^2)/(2 \Delta_a)$, which is explicitly calculated in the SM~\cite{suppMat}.
In order to study the generation of entanglement in the presence of decoherence, we calculate the Dicke spin squeezing parameter~\cite{VacuumSqueezing};
\begin{equation}\label{eq:SqueezeParam}
\xi^2 = \min_\theta \frac{\text{var}(\hat{J}^\perp_\theta) }{|\langle \textbf{J} \rangle|/2},
\end{equation}
where $\text{var}(\hat{A}) = \langle\hat{A}^2\rangle_\psi -\langle\hat{A}\rangle_\psi^2$ is the variance, $\hat{J}^\perp_\theta$ is the pseudo-spin tilted at angle $\theta$ and perpendicular to the average spin length, $\mathbf{J}$.
Notably, $\Delta\phi^2 = \xi^2/N$ and therefore, $\xi^2 < 1$ indicates better than SQL certainty and the presence of metrologically useful entanglement~\cite{OptimalGenerators_Wilson}.

We assume that the squeezing is done for short times, such that $N\Gamma_+ t$, $N\Gamma_+ t\ll1$ and collective decoherence doesn't drive atoms out of the two-level manifold.
This assumption allows dissipative dynamics to be treated independently of the OAT dynamics, so we add uncertainty due to collective emission in quadrature to variance of the squeezed state.
We calculate the mean-field dynamics~\cite{suppMat}, where we find
\begin{equation}
\xi^2 = \frac{1}{(N \chi t/2)^2} + t \gamma_l + N t \Gamma_c,
\end{equation}
where $\Gamma_C = 2 \left( \Gamma_+ + \Gamma_- \right) $ is the contribution due to collective decoherence.
In Fig.~\ref{fig:Fig4} (a) we show these four rates, for $N=3000$.
The spin squeezing parameter is minimized as a function of squeezing time, $t$, and detuning of the pump from the dressed cavity detuning, $\Delta_c'$.

In order to achieve quantum enhanced sensitivity, we consider the far detuned regime where collective decay is on the order of single particle effects.
By balancing the single particle decoherence and collective decoherence, we find $\Delta_c'$ such that $N \Gamma_c=\gamma_l$, shown in Fig.~\ref{fig:Fig4}(a) and (b) by the green dashed vertical line.
To highest order in $N$, we find an optimal squeezing time and dressed cavity detuning of
\begin{equation}
t_\mathrm{opt} \approx 4 \left( \frac{2 N^2 \Delta_a^5 \kappa^4}{ U_0 \gamma^5 \eta^6 } \right)^{1/3}, \quad \Delta'_{c,\mathrm{opt}} \approx \pm  \sqrt{ \frac{2 N U_0 \Delta_a \kappa}{\gamma} }
\end{equation}
and an optimal spin squeezed parameter of
\begin{equation}
\xi^2_\mathrm{opt} \approx 3 \left( \frac{2 \gamma \kappa }{N U_0 \Delta_a} \right)^{1/3} ,
\end{equation}
which corresponds to $\Delta\phi^2 \propto 1/N^{4/3}$ and is better than SQL scaling. 
Lastly, if using the fact that $\Delta_a^2 + \gamma^2/4 \approx \Delta_a^2$, we have $U_0 = g^2 /(2 \Delta_a) $, and
\begin{equation}
\begin{aligned}
t_\mathrm{opt}& \approx \frac{8 \kappa \Delta_a^2}{ \gamma^2 \eta^2 } \left( \frac{2 N^2 }{\mathcal{C}_1} \right)^{1/3}, \Delta'_{c,\mathrm{opt}} \approx \pm  \frac{\kappa}{2} \sqrt{ N \mathcal{C}_1 },\\
& \quad \quad \quad \xi^2_\mathrm{opt} \approx 3 \left( N \mathcal{C}_1 \right)^{-1/3} ,
\end{aligned}
\end{equation}
where $\mathcal{C}_1 \equiv 4 g^2 / ( \gamma \kappa )$ is an effective single atom co-operativity.
In Fig.~\ref{fig:Fig4} (b), we show that for an atomic detuning of $2 \pi \times 150$ MHz, and a pump rate of $\eta = 2\pi \times 10 $ MHz, the optimally squeezed state is reached in a time of 150 $\mu$s.

\emph{Conclusion}. ---
In this work, we provided a detailed derivation of the momentum exchange interaction in Ref.~\cite{MomentumExchange_Luo}, and discussed its connection to the frequency shift of an optical cavity mode due to atomic motion.
We have introduced a method which may be used to realize momentum based entanglement independent of electronic degrees of freedom with many possible squeezing interactions, shown in Eq.~\eqref{eq:Hatom}.
These squeezing interactions are generated from the cavity field's response to the atomic motion.
We showed that, in proper limits, the momentum states may be truncated to form an effective two-level system to the momentum states $0 \hbar k - p_0$ and $+2 \hbar k - p_0$,  which form a density grating inside the cavity.
Using this effective two-level system, we have demonstrated that entanglement is generated even in the presence of noise.

Here, we've limited the discussion of entanglement generation to an effective two-level truncation, but one could go further to realize multi-level squeezing dynamics from Eq.~\eqref{eq:Hatom}.
Of particular interest is the use of these additional states to realize multi-level dynamics via the use of additional dressing lasers to drive higher order interactions.
Furthermore, these momentum states provide a promising opportunity for realizing spin squeezing via two-axis counter twisting effects, which can be achieved by periodically modulating the amplitude injected field at a parametric resonance~\cite{PDD_Reilly} or injecting a second pump~\cite{TACT_luo}.
Future work could include using measurements on the light field exiting the cavity.
This light is highly entangled to the atoms, and measurements would, in principle, allow one to induce a back-action on the atomic cloud.
This creates the opportunity for quantum non-demolition measurements similar to those presented in Ref.~\cite{EntangledMomentum_Greve}, or even for extending methods of continuous phase tracking~\cite{ContinuousPhase_Shankar} directly to matter-wave interferometry.

\emph{Acknowledgments}. ---
The authors thank Simon B. J\"ager, Maya Miklos, Tyler Sterling, and John Cooper for helpful conversations.
This material is based upon work supported by the U.S. Department of Energy, Office of Science, National Quantum Information Science Research Centers, Quantum Systems Accelerator. We acknowledge additional funding support from the National Science Foundation under Grant Numbers PHY 2317149; OMA 2016244; PHY 2317149; PHY 2207963; and 2231377, NIST, and DARPA/ARO W911NF-19-1-0210; and W911NF-16-1-0576, AFOSR grants FA9550-18-1-0319; and FA9550-19-1-0275.

\providecommand{\noopsort}[1]{}\providecommand{\singleletter}[1]{#1}%

\pagebreak

\clearpage

\appendix

\section{Numerical Simulation Parameters}\label{sec:params}
For the two relevant atomic levels we consider the $D_2$ cycling transition, with a transition wavelength of $780$ nm and a decay rate of $\gamma = 2 \pi \times 6.066$ MHz~\cite{Rb_Steck}. 
The single relevant mode of the cavity couples to the atoms with a single-atom coupling strength of $g=2\pi \times 0.5$ MHz and decays at a rate $\kappa = 2\pi \times 0.055$ MHz.
The atom-pump detuning is $\Delta_a = 2\pi \times 150$ MHz with a pump coupling rate of $\eta = 2\pi \times 10 $ MHz, corresponding to the rate at which pump photons make it through the mirror.
The atoms are accelerated to have a mean momentum of $p_0 = 15 \hbar k$, which provides an energy splitting of $\omega_z = - 2 \pi \times 0.44$ MHz, which matches Ref.~\cite{MomentumExchange_Luo} to an order of magnitude.

\section{The Dipole radiation Pattern} 
The dipole radiation pattern is calculated by considering a specific dipole radiation pattern.
As an example, in simulations we consider the rubidium dipole allowed transition from $F=2, m_F = 2$ to $F=3,m_f = 3$ and tracing out the $x$ and $y$ degrees of freedom~\cite{Doppler_Castin,Thesis_Bartolotta}.
This yields 
\begin{equation}
\begin{aligned}
\hat{d}_\mu(\theta) = \sqrt{\mathcal{N}(\theta)} \exp[-i k \hat{z}_\mu \cos(\theta)], \quad \mathcal{N}(\theta) =\frac{3}{8}
(1+\cos(\theta)^2),
\end{aligned}
\end{equation}
where $\theta$ is the angle between the emitted photon and the positive $z$-axis~\cite{Doppler_Castin,Thesis_Bartolotta}.
If we had considered states with another angular momentum difference, this would lead to a different dipole radiation pattern.
The full term in the Lindblad equation is then integrated over $\theta$:
\begin{equation}\label{eq:spontemiss}
\begin{aligned}
\frac{\hat{\mathcal{D}}[\hat{L}_{\mathrm{sp},\mu}] \hat{\rho}}{\gamma} = \int_{-1}^{+1} \bigg[ &\mathcal{N}(u) \exp(-i k \hat{z}_\mu u) \hat{\sigma}^-_\mu \hat{\rho} \hat{\sigma}^+_\mu \exp(+i k \hat{z}_\mu u) \\
& - \{ \hat{\sigma}^+_\mu \hat{\sigma}^-_\mu, \hat{\rho} \}/2 \bigg] du.
\end{aligned}
\end{equation}

\section{Adiabatic Elimination of the Excited State}\label{sec:AdiabaticElim}
From Eq. (1) in the main text, we eliminate the excited state based on the assumption $ \abs{\Delta_a} \gg \sqrt{N} g$, with $\Delta_a \equiv \omega_a - \omega_p$. 
We assume that all atoms start in the ground state, i.e. $\hat{\rho}_\mathrm{org} = \hat{\rho}_{k,c} \otimes \hat{\rho}_a$ where
\begin{equation}\label{eq:rhoa}
\hat{\rho}_a = \bigotimes^N_{\mu=1} \ket{g}_\mu\bra{g}_\mu .
\end{equation}
Now, we because the excited state is far detuned, any frequencies (timescales) are much bigger (faster) for excited state dynamics compared to the relevant atomic motion we wish to study.
Subject to this assumptions, we take make the approximation that $(\partial/\partial t) \bra{e}_\mu \hat{\rho}_a \ket{g}_\mu \approx 0$ for all relevant timescales.
Now, using the Heisenberg picture of evolution:
\begin{equation}
\frac{\partial \hat{A}}{\partial t} = \frac{i}{\hbar} \left[ \hat{H}, \hat{A} \right] + \sum_j \hat{\mathcal{D}}^\dagger [\hat{L}_j] \hat{A}.
\end{equation}
this approximation is equivalent to the condition that
\begin{equation}
(\partial/\partial t) \hat{\sigma}^-_\mu\approx0.
\end{equation}
Therefore, we can directly solve for this condition in order to find an effective replacement for the atom internal state observables:
\begin{equation}
(\partial/\partial t) \hat{\sigma}^-_\mu = - i \Delta_{a} \hat{\sigma}^-_\mu + i g \cos(k \hat{z}_\mu) \hat{a} \hat{\sigma}^z_\mu - \frac{\gamma}{2} \hat{\sigma}^-_\mu \approx 0,
\end{equation}
and therefore
\begin{equation}
\begin{aligned}
\hat{\sigma}^-_\mu &\approx \frac{g}{2 \Delta_{a} - i \gamma } \cos(k \hat{z}_\mu) \hat{a} \hat{\sigma}^z_\mu.
\end{aligned}
\end{equation}
Now, because we have assumed Eq.~\ref{eq:rhoa}, we may replace $\hat{\sigma}^z_\mu \rightarrow - \ket{g}_\mu\bra{g}_\mu $, where the atoms are always in the ground state so $\ket{g}_\mu\bra{g}_\mu = 1$, yielding
\begin{equation}
\begin{aligned} \label{eq:sigmaMinus}
\hat{\sigma}^-_\mu &\approx \frac{-g}{2 \Delta_{a} - i \gamma } \cos(k \hat{z}_\mu) \hat{a}.
\end{aligned}
\end{equation}

To reconstruct the correct Hamiltonian and jump operators, we consider the time evolution of the cavity operator $\hat{a}$, and of the momentum operator $\hat{p}_\mu$:
\begin{equation}
\begin{aligned}
\frac{\partial}{\partial t} \hat{a} = & \frac{i}{\hbar} [\hat{H}_\mathrm{rot},\hat{a}] +\hat{\mathcal{D}}^\dagger[\hat{L}_\mathrm{cav}] \hat{a} \\
\frac{\partial}{\partial t} \hat{p}_\mu = & \frac{i}{\hbar} [\hat{H}_\mathrm{rot},\hat{p}_\mu] + \hat{\mathcal{D}}^\dagger[\hat{L}_{\mathrm{sp},\mu}] \hat{p}_\mu
\end{aligned}
\end{equation}
where $\hat{\mathcal{D}}^\dagger[\hat{L}_{\mathrm{sp},\mu}] \hat{p}_\mu$ includes integration about $\theta$ as in Eq.~\eqref{eq:spontemiss}.
We replace the action of $\hat{\sigma}^-_\mu$ with the right hand side of Eq.~\ref{eq:sigmaMinus} to find the atom-field Hamiltonian and jump operators
\begin{equation}
\begin{aligned}
\hat{H}_\mathrm{af} = &  \hbar \Delta'_{c} \hat{a}^\dagger\hat{a} + \hbar \eta \left( \hat{a} + \hat{a}^\dagger \right) \\
& + \sum_{\mu=1}^N \left( \frac{\left( \hat{p}_\mu- p_0 \right)^2}{2m} - \hbar U_0 \  \hat{a}^\dagger \hat{a} \cos(2 k \hat{z}_\mu) \right), \\
\hat{L}_\mathrm{cav} = & \sqrt{\kappa} \hat{a}, \quad \hat{L}_{\mathrm{rec},\mu}(\theta) = \sqrt{2 \gamma U_0/\Delta_a } \ \hat{d}_\mu(\theta) \hat{a} \cos( k \hat{z}_\mu ),
\end{aligned}
\end{equation}
with $\Delta'_c = \Delta_c - U_0 N$ being the dressed cavity detuning.
The state may now be found by 
\begin{equation}
\Tr_a( \hat{\rho}_\mathrm{org} ) = \hat{\rho}_{k,c} \Tr_a(\hat{\rho}_a) = \hat{\rho}_{k,c},
\end{equation}
where we have again used the assumption that Eq.~\ref{eq:rhoa} is true for all relevant timescales.

\section{Discretized Momentum Space} \label{sec:DiscMomentum}

We note that the single particle jump operator does not fit conveniently in this formalism, but later when considering squeezing we treat single particle decoherence explicitly.
The entire Hamiltonian now contains only collective momentum physics, which we second quantize via
\begin{equation}
\begin{aligned}
\int \hat{\psi}^\dagger(p) \hat{\psi}(p) dp =& \\
\sum_{n} \int_{(2 n -1 )\hbar k}^{(2 n + 1) \hbar k} \hat{\psi}^\dagger(p) \hat{\psi}(p) dp =&  \\
\sum_{n} \hat{b}^\dagger_n \hat{b}_n,
\end{aligned}
\end{equation}
so that each sum over all particles is replaced by an integral over an operator valued measure.
We also impose that $\sum_{n} \hat{b}^\dagger_n \hat{b}_n = N$, ie there are $N$ total atoms.
We additionally assume the particles have near zero-momentum width, so that
\begin{equation}
\begin{aligned}
\hat{b}_n^\dagger \hat{b}_n =& \int_{(2 n -1 )\hbar k}^{(2 n + 1) \hbar k} \hat{\psi}^\dagger(p) \hat{\psi}(p) dp \\
\approx& \int_{(2 n -\epsilon )\hbar k}^{(2 n + \epsilon) \hbar k} \hat{\psi}^\dagger(p) \hat{\psi}(p) dp,
\end{aligned}
\end{equation}
for $\epsilon \ll 1$.
This means that $2 n \hbar k + p \approx 2 n \hbar k$ in terms like the kinetic energy.
Going term by term, this yields;
\begin{equation}
\begin{aligned}
\sum_j \frac{\left( \hat{p}_j - p_0\right)^2}{2 \hbar m} &= \int_{-\infty}^{+\infty} \hat{\psi}^\dagger(p) \frac{\left( p - p_0\right)^2}{2 \hbar m} \hat{\psi}(p) dp \\
& \approx \sum_{n} \int_{(2 n -\epsilon )\hbar k}^{(2 n + \epsilon) \hbar k} \hat{\psi}^\dagger(p) \frac{\left( 2 n \hbar k - p_0\right)^2}{2 \hbar m} \hat{\psi}(p) dp \\
& = \sum_{n} \hat{b}_n^\dagger \hat{b}_n K_n,
\end{aligned}
\end{equation}
where $K_n = \left[ (n-n_0) 2 \hbar k \right]^2/2\hbar m $ is the kinetic energy for the $n^\mathrm{th}$ momentum state and $n_0$ is the mean momentum state such that $p_0 = n_0 2 \hbar k$.
Each of the trigonometric operators is simply expressed in this formalism as well;
\begin{equation}
\begin{aligned}
\sum_j \mathrm{e}^{\pm i 2 k \hat{z}_j} &= \sum_{n} \int_{(2 n -1 )\hbar k}^{(2 n + 1) \hbar k} \hat{\psi}^\dagger(p) \mathrm{e}^{\pm i 2 k \hat{z}} \hat{\psi}(p) dp \\
& = \sum_{n} \int_{(2 n -1 )\hbar k}^{(2 n + 1) \hbar k} \hat{\psi}^\dagger(p \pm 2 \hbar k) \hat{\psi}(p)dp \\
& = \sum_{n} \hat{b}_{n\pm1}^\dagger \hat{b}_n \equiv \sum_{n} \hat{J}^\pm_{n},
\end{aligned}
\end{equation}
which we use to write
\begin{equation}
\begin{aligned}
\hat{C}_n \equiv& \frac{\hat{J}^+_{n} +\hat{J}^-_{n+1}}{2}, \quad \sum_j \cos(2 k \hat{z}_j) = \sum_n \hat{C}_n,
\end{aligned}
\end{equation}
and
\begin{equation}
\begin{aligned}
\hat{S}_n \equiv& \frac{\hat{J}^+_{n} -\hat{J}^-_{n+1}}{2i}, \quad \sum_j \sin(2 k \hat{z}_j) = \sum_n \hat{S}_n .
\end{aligned}
\end{equation}
We note that $[\hat{C}_n, \hat{S}_n] = \frac{i}{2} ( \hat{b}_{n+1}^\dagger \hat{b}_{n+1} - \hat{b}_n^\dagger \hat{b}_n )$, following from an $\mathfrak{su}(2)$ commutation relation.
This means the Hamiltonian after displacement is
\begin{equation}
\begin{aligned}
\frac{\tilde{\hat{H}}_\mathrm{af}}{\hbar} =& \Delta_c' \tilde{a}^\dagger \tilde{a} + \sum_n \left( K_n \hat{b}^\dagger_n \hat{b}_n  - U_0 \abs{\beta}^2 \hat{C}_n \right) \\
&- U_0 \left( \tilde{a}^{\dagger} \tilde{a} + \tilde{a}^{\dagger} \beta + \beta^* \tilde{a} \right) \sum_n \hat{C}_n .
\end{aligned}
\end{equation}

\begin{widetext}

\section{scattered photons in the cavity} \label{sec:a_elim}
Now, we eliminate the cavity degrees of freedom from Eq.~(3).
As noted in the main text, the cavity response to the pump and atomic motion can be separated into two components.
The pump leads to a displacement of the cavity mode by amount $\beta \equiv \eta/(-\Delta_c' + i \kappa/2)$, yielding
\begin{equation}
\hat{D}_1(\beta)^\dagger \hat{a} \hat{D}_1(\beta) = \beta + \hat{a},
\end{equation}
where $\hat{D}_1(\beta) \equiv \exp( \hat{a}^\dagger \beta - \beta^* \hat{a} )$ is the displacement operator.
Now $\beta$ is the coherent injected field amplitude.
The Hamiltonian in this displaced picture is correspondingly
\begin{equation}
\begin{aligned}
\frac{\tilde{H}_\mathrm{af}}{\hbar} \equiv& \Delta_c' \hat{a}^\dagger \hat{a} + \Delta_c' |\beta|^2 + \sum_{n} K_n \hat{b}_n^\dagger \hat{b}_n - U_0 \hat{C}_n ( \beta^* + \hat{a}^\dagger )( \beta + \hat{a} ) \\
=& ( \Delta_c' - U_0 \sum_n \hat{C}_n ) \hat{a}^\dagger \hat{a} + \Delta_c' |\beta|^2 + \sum_{n} K_n \hat{b}_n^\dagger \hat{b}_n - U_0 |\beta|^2 \hat{C}_n - U_0 \beta^* \hat{C}_n \hat{a} - U_0 \beta \hat{C}_n \hat{a}^\dagger. \\
\end{aligned}
\end{equation}
The energy due to the classical field, $\Delta_c' |\beta|^2$ is just a constant real number so it will only add a global phase and we can drop it.
Now, the Hamiltonian has three components.
First, $\tilde{H}_k \equiv \hbar \sum_n \left( K_n \hat{b}^\dagger_n \hat{b}_n  - U_0 \abs{\beta}^2 \hat{C}_n \right)$ is the momentum only Hamiltonian, where now the classical injected field drives a cosine interaction with the atoms.
Second, $\tilde{H}_c = \hbar \Delta_c' \tilde{a}^\dagger \tilde{a}$ is the photon only Hamiltonian.
Lastly, the interaction Hamiltonian is $\tilde{H}_{\mathrm{int}} = -\hbar U_0 \sum_n \hat{C}_n ( \hat{a}^\dagger \hat{a} - \beta^* \hat{a} - \beta \hat{a}^\dagger ) $,
where the atomic motion modifies the frequency of the photons through the first term, and scatters photons in to and out of the coherent field through the second and third terms.
Now, these photons start in a displaced vacuum about $\beta$ and, as time goes on, only acquire non-trivial dynamics through the coupling to atomic motion, coupled by a small frequency $U_0$. 
Therefore, we can adiabatically elminate the photons.
To do this, we use one more displacement
\begin{equation}
\hat{D}_2(\hat{\alpha}(t)) \equiv \exp( \tilde{a}^\dagger \hat{\alpha}(t) - \hat{\alpha}(t)^\dagger \tilde{a} )
\end{equation}
where $\hat{\alpha}(t)$ is a position/momentum operator, and we solve for the new master equation.
This displacement replaces the photonic equations of motion with the equations of motion for the operators that drive them, in this case $\hat{J}_n^\pm$ representing the two photon recoil.
We solve for the equations of motion of $\hat{\alpha}(t)$ following Ref.~\cite{Lindblad_jager};
\begin{equation}
\begin{aligned}
\frac{\partial}{\partial t} \hat{\alpha}(t) =& - \frac{i}{\hbar} \left[ \hat{H}_k, \hat{\alpha}(t) \right] - i \left( \Delta_c' - U_0 \sum_n \hat{C}_n \right) \hat{\alpha}(t) - \frac{\kappa}{2} \hat{\alpha}(t) + i U_0 \beta \sum_n \hat{C}_n.
\end{aligned}
\end{equation}
Here, we observe that $\Delta_c' - U_0 \sum_n \hat{C}_n = \Delta_c'( 1 - \frac{U_0}{\Delta_c'} \sum_n \hat{C}_n ) \approx \Delta_c'$.
Dropping this term is equivalent to dropping the next order of modulation, i.e. the frequency shift of the quantum degrees of freedom $\hat{\alpha}(t)$ by the atoms.
In other words, we only solve for the central cavity peak interacting with the atoms (first order), the frequency modulation of the central cavity peak (second order), and the interaction of the atoms with that peak (third order), and not the fourth order modulation of these scattered photons--which is the same order as the modulation due to spontaneous emission that we already dropped.
All together, we must solve
\begin{equation}
\begin{aligned}
\frac{\partial}{\partial t} \hat{\alpha}(t) =& - i \left[ \sum_n K_n \hat{b}^\dagger_n \hat{b}_n, \hat{\alpha}(t) \right] - i \left( \Delta_c' - i \frac{\kappa}{2} \right) \hat{\alpha}(t) + i U_0 \beta \sum_n \hat{C}_n.
\end{aligned}
\end{equation}
We can make the ansatz that
\begin{equation}
\hat{\alpha}(t) = \beta \sum_n \left( A^+_n(t) \hat{J}^+_n + A^-_n(t) \hat{J}^-_n \right)/2,
\end{equation}
where the amplitude $A^{\pm}_n(t)$ carries the time and momentum dependence for photon absorption in the $\pm z$ direction, and $\hat{J}^\pm_n$ is the corresponding momentum kick direction in that direction.
We note that
\begin{equation}
\begin{aligned}
\left[ \sum_n  K_n\hat{b}^\dagger_n \hat{b}_n, \sum_m \hat{J}^\pm_m \right] = \sum_n \left( K_{n\pm1} - K_n \right) \hat{J}^\pm_n.
\end{aligned}
\end{equation}
This ansatz yields two uncoupled differential equations;
\begin{equation}
\begin{aligned}
\dot{A}^\pm_n(t) = -i ( K_{n\pm1} - K_n ) \ A^\pm_n(t) - i \left( \Delta_c' - i \frac{\kappa}{2} \right) A^\pm_n(t) + i U_0,
\end{aligned}
\end{equation}
which have solutions
\begin{equation}
A^\pm_n \approx \frac{U_0}{\Delta_c' + (K_{n\pm1} - K_n) - i \kappa/2},
\end{equation}
where we've dropped the initial time dependent decay from the pump being turned on.
Now, we note that $[ \hat{\alpha}, \sum_n \hat{C}_n ] = 0$ and substitute $\hat{\alpha}$ this back into the Hamiltonian;
\begin{equation}
\begin{aligned}
\frac{\hat{H}_a}{\hbar} =& \sum_{n} \left( K_n \hat{b}_n^\dagger \hat{b}_n - U_0 |\beta|^2 \hat{C}_n - \frac{U_0}{2} \left[ \beta \hat{\alpha}^\dagger + \beta^* \hat{\alpha} \right] \hat{C}_n \right) \\
=& \sum_{n} \left( K_n \hat{b}_n^\dagger \hat{b}_n - U_0 |\beta|^2 \hat{C}_n - \frac{U_0|\beta|^2}{2} \sum_m \mathfrak{R}\left[A^+_m \hat{J}^+_m + A^-_m \hat{J}^-_m\right] \hat{C}_n \right).
\end{aligned}
\end{equation}
Now, the atoms interacts with the injected field, of intensity $|\beta|^2$, and the atoms feel the effects of their own motion reflected back on them through this second term.
In order to make physical sense of this second term, we write $\hat{J}^\pm_m$ in the real (hermitian) and imaginary (anti-hermitian) components:
\begin{equation}
\begin{aligned}
\hat{J}^+_m =& \hat{C}_m + i \hat{S}_m, \quad \mathrm{and} \quad \hat{J}^-_m = \hat{C}_{m-1} + i \hat{S}_{m-1}  .
\end{aligned}
\end{equation}
We also note that $\mathfrak{R}(Y Z) = \mathfrak{R}(Y) \mathfrak{R}(Z) - \mathfrak{I}(Y) \mathfrak{I}(Z)$.
Therefore
\begin{equation}
\begin{aligned}
\sum_m \mathfrak{R}\left[A^+_m \hat{J}^+_m + A^-_m \hat{J}^-_m\right] =& \sum_m \left[ \mathfrak{R}\left( A^+_m\right) \hat{C}_m + \mathfrak{R}\left( A^-_m \right)  \hat{C}_{m-1} - \mathfrak{I}\left( A^+_m\right) \hat{S}_{m} - \mathfrak{I}\left( A^-_m \right)  \hat{S}_{m-1}  \right] \\
=& \sum_m \left[ \mathfrak{R}\left( A^+_m + A^-_{m+1} \right) \hat{C}_{m} - \mathfrak{I}\left( A^+_m - A^-_{m+1} \right) \hat{S}_{m} \right].
\end{aligned}
\end{equation}
This yields
\begin{equation}
\begin{aligned}
\frac{\hat{H}_a}{\hbar} =& \sum_{n} \left( K_n \hat{b}_n^\dagger \hat{b}_n - U_0 |\beta|^2 \hat{C}_n - \sum_m \left[ \chi_m \hat{C}_m + \zeta_m \hat{S}_m \right] \hat{C}_n \right).
\end{aligned}
\end{equation}
where 
\begin{equation}
\begin{aligned}
\chi_n &\equiv  \frac{U_0 |\beta|^2 }{2} \mathfrak{R}( A^+_n + A^-_{n+1} ),\ \zeta_n \equiv  \frac{U_0 |\beta|^2 }{2} \mathfrak{I}( A^-_{n+1} - A^+_n ) 
\end{aligned}
\end{equation}
are the rates for the cosine and sine non-linear interactions.
This adiabatic elimination also yields the collective jump operator $ \hat{L}_c = \sqrt{ \kappa |\beta|^2 } \ \sum_{n} \left( A^+_n \hat{J}^+_n + A^-_n \hat{J}^-_n\right)/2$.

\end{widetext}

\section{Two Level Mapping}\label{sec:2level}

\begin{figure*}

\centerline{\includegraphics[width=0.8 \textwidth]{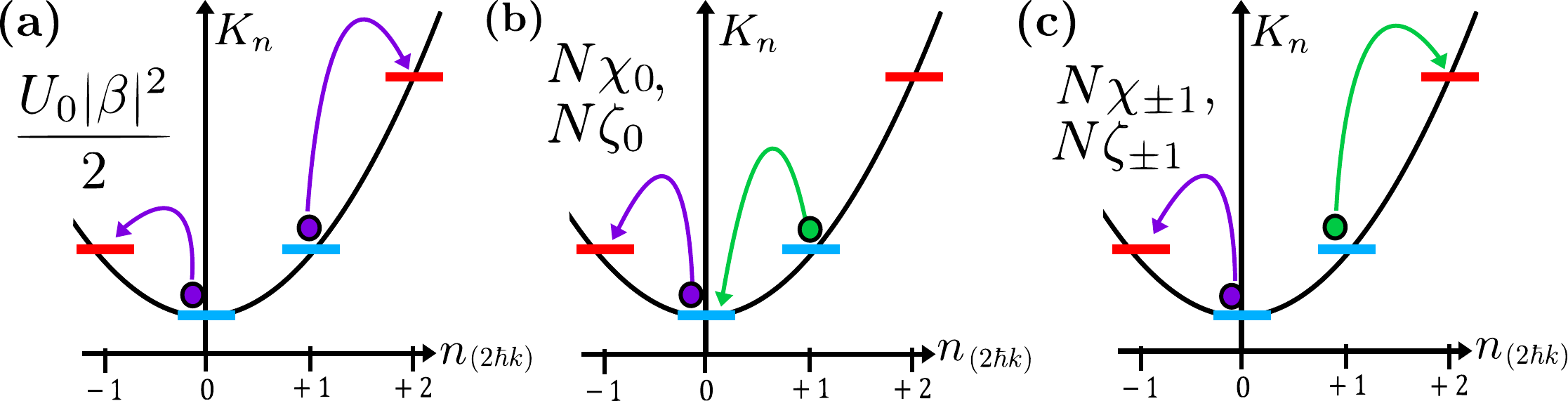}}
    \caption{(a) Single atom processes which break the two-level assumption, due to Bragg scattering off of the optical lattice.
    (b) and (c) Multi-atom process which break the two-level approximation.
    (b) An example transition which occurs with rate $N \chi_0$ or $N \zeta_0$, where one atom (purple) leaves the two-level manifold but the other atom (green) does not.
    This is the multi-atom process which pre-dominantly breaks the two-level approximation.
    (c) An example transition which occurs with rate $N \chi_1$ or $N \zeta_1$, where both atom (purple and green) leave the two-level manifold.
    This the multi-atom process has a higher energy barrier than if one atom stayed in the manifold.
    }
    \label{fig:Hopping}
\end{figure*}

Now, we restrict ourselves to just $n=0,+1$ (the $0 \hbar k$ and $+2 \hbar k$ states).
By restricting to these two levels, we are left with only two-level operators;
\begin{equation}
\begin{aligned}
\hat{J}_x \equiv& \hat{C}_0 = \frac{1}{2} \left( \hat{J}^+_0 + \hat{J}^-_1 \right),\\
\hat{J}_y \equiv& \hat{S}_0 = \frac{1}{2i} \left( \hat{J}^+_0 - \hat{J}^-_1 \right) \\
\hat{J}_z \equiv& -i [ \hat{J}_x, \hat{J}_y ] = \frac{1}{2} ( \hat{b}^\dagger_1 \hat{b}_1 - \hat{b}^\dagger_0 \hat{b}_0 )\\
\end{aligned}
\end{equation}
These two momentum states are separated by a kinetic energy $ \hbar \omega_z \equiv \hbar ( K_{+1} - K_0 ) $.

Now, we use the following:
\begin{equation}
\hat{b}_n \ket{\psi} = 0, \quad \mathrm{and} \quad \hat{J}^\pm_n \ket{\psi} = 0,\quad \mathrm{for} \quad n\neq 0,+1.
\end{equation}
This means
\begin{equation}
\begin{aligned}
\frac{\hat{H}_a}{\hbar} =& K_0 \hat{b}_0^\dagger \hat{b}_0 + K_1 \hat{b}_1^\dagger \hat{b}_1 - U_0 |\beta|^2 \hat{C}_0  \\
& - \chi_0 \hat{C}_0^2 + \zeta_0 ( \hat{S}_0 \hat{C}_0 + \hat{C}_0 \hat{S}_0 ) \\
& - \hat{B}_1 -  \hat{B}_2 - \hat{B}_3 - \hat{B}_4
\end{aligned}
\end{equation}
where the terms labeled $\hat{B}_j$ for $j = 1, 2, 3, 4$ break the two-level approximation;
\begin{equation}
\begin{aligned}
\hat{B}_1 \equiv& U_0 |\beta|^2 \left( \hat{J}^-_0 + \hat{J}^+_1 \right)/2\\
\hat{B}_2 \equiv& \left[ \chi_0 \hat{C}_0 + \zeta_0 \hat{S}_0 \right] \left( \hat{J}^-_0 + \hat{J}^+_1  \right)/2 \\
\hat{B}_3 \equiv& \left[ \chi_{-1} \hat{J}^-_{0} + \chi_{1} \hat{J}^+_{1} + \zeta_{-1} \hat{J}^{-}_0 + \zeta_1 \hat{J}^+_1  \right] \hat{C}_0 \\
\hat{B}_4 \equiv& \left[ \chi_{-1} \hat{J}^-_{0} + \chi_{1} \hat{J}^+_{1} + \zeta_{-1} \hat{J}^{-}_0 + \zeta_1 \hat{J}^+_1  \right] (\hat{J}^-_0 + \hat{J}^+_1 )/2
\end{aligned}
\end{equation}
if we only keep the terms that keep us in the two-level manifold we get, and gauge transform away terms proportional to $ ( \hat{b}_0^\dagger \hat{b}_0 + \hat{b}_1^\dagger \hat{b}_1) $, then we find
\begin{equation}
\begin{aligned}
\frac{\hat{H}_{\mathrm{eff}}}{\hbar} =& \omega_z \hat{J}_z - U_0 |\beta|^2 \hat{J}_x - \chi_0 \hat{J}_x^2 - \frac{\zeta_0}{2} \left( \hat{J}_x \hat{J}_y + \hat{J}_y \hat{J}_x \right), \\
& \hat{L}_c = \sqrt{\kappa |\beta|^2 / 4} \ ( A^+_0 \hat{J}^+ + A^-_{+1} \hat{J}^- ),
\end{aligned}
\end{equation}
where we drop the subscript in the main text: $\chi_0 \rightarrow \chi$ and $\zeta_0 \rightarrow \zeta$.

\subsection{Restrictions on the Two-Level Mapping}\label{sec:Restrictions}

Separating these two momentum states requires that none of the terms $\hat{B}_j$ drive atoms out of this two level manifold.
These terms amount to Bragg scattering, $\hat{B}_1$, and the nonlinear cosine-cosine and cosine-sine interactions, $\hat{B}_2$, $\hat{B}_3$, and $\hat{B}_4$.

The first term, Bragg scattering, could break the two level approximation by allowing the following single particle transition:
\begin{equation}
\ket{0 \hbar k} \rightarrow \ket{-2 \hbar k} \quad \mathrm{or} \quad \ket{+2 \hbar k} \rightarrow \ket{+4 \hbar k}
\end{equation}
through the transition elements; 
\begin{equation}
\begin{aligned}
\bra{n=-1} \hat{B}_1 \ket{n=0} =& \frac{U_0 |\beta|^2}{2} \\
\bra{n=+2} \hat{B}_1 \ket{n=+1} =& \frac{U_0 |\beta|^2}{2},
\end{aligned}
\end{equation}
shown schematically in Fig.~\ref{fig:Hopping} (a).
These single particle terms come with the energy condition:
\begin{equation}
| K_{-2} - K_0 |, \quad \mathrm{and} \quad | K_{+2} - K_{+1} |.
\end{equation}
Using this, we can consider the effective Rabi-oscillation between $0\hbar k$ and $-2 \hbar k$ or $+2\hbar k$ and $+4 \hbar k$, which yields the energy condition the maximum probability of excitation~\cite{QO_Steck}:
\begin{equation}
\frac{ \left( \frac{U_0 |\beta|^2}{2} \right)^2 }{\left( \frac{U_0 |\beta|^2}{2} \right)^2 + |\omega_k|^2 } \ll 1
\end{equation}
for $\omega_k = \min\left( | K_{-1} - K_0| ,\  | K_{+2} - K_{+1} | \right)$ being the smaller of the two energy gaps.
This yields the condition that
\begin{equation}
U_0 |\beta|^2 \ll 2 \min\left(\ | K_{-1} - K_0| ,\  | K_{+2} - K_{+1} |\ \right).
\end{equation}
In the limit that that atoms are fast moving, $p_0 \gg 2 \hbar k$, this yields
\begin{equation}
U_0 |\beta|^2 \ll \frac{4 k p_0}{m}.
\end{equation}
which means that the standing field will also have low probability of driving $\ket{0 \hbar k} \leftrightarrow \ket{+2 \hbar k}$ as well.

The second set of terms, $\hat{B}_2$, $\hat{B}_3$, and $\hat{B}_4$ correspond to non-linear interactions.
These drive two atom transitions, such as the two examples in Fig.~\ref{fig:Hopping} (b) and (c) and therefore have many more ways to break the two-level approximation. 
For two atomic states labeled by the pair $(n,m) = \mathcal{S}( \ket{n}\otimes\ket{m})$ with the symmetrizer $\mathcal{S}$ and momentum states $\ket{n} = \ket{n 2 \hbar k}$, the transitions which break the two level approximation are
\begin{equation}
\begin{aligned}
&(0,0) \rightarrow (-1,-1) && 2 ( K_{-1} - K_0 ) \\
&(0,0) \rightarrow (-1,+1) && K_{+1} + K_{-1} - 2 K_0 \\
    &(0,+1) \rightarrow (-1,0) && K_{-1} - K_{+1} \\
    &(0,+1) \rightarrow (-1,+2) && K_{-1} + K_{+2} - K_{0} - K_{+1} \\
    &(0,+1) \rightarrow (+1,+2) && K_{+2} - K_{0} \\
    &(+1,+1) \rightarrow (0,+2) && K_{0} + K_{+2} - 2 K_{+1} \\
    &(+1,+1) \rightarrow (+2,+2) && 2 ( K_{+2} - K_{+1} ) \\
\end{aligned}
\end{equation}
Each of these have a corresponding matrix element, such as in the $\hat{B}_1$ case.
However, the transitions $(0,0)\rightarrow(-2,+2)$ and $(2,2)\rightarrow(0,+4)$ from will always have the lowest energy cost with $K_{+1} + K_{-1} - 2 K_0 = K_{0} + K_{+2} - 2 K_{+1} = 8 \hbar^2 k^2 / 2 \hbar m$ . 
Furthermore, we care about maximizing $\chi_0$ and $\zeta_0$ to optimize squeezing on this two level manifold, so we only bound these two rates.
In principle, this process provides bounds for every rate appearing in $\hat{B}_j$.
This gives:
\begin{equation}
N \abs{\chi_0},\ N \abs{\zeta_0} \ll \frac{ 8 \hbar^2 k^2}{2 \hbar m}
\end{equation}

So, these two restrictions together are 
\begin{equation} 
\begin{aligned}
\abs{U_0} \abs{\beta}^2 \ll \frac{4 k p_0}{m}, \quad \mathrm{and} \quad N \abs{\chi_0},\ N \abs{\zeta_0} \ll \frac{ 8 \hbar^2 k^2}{2 \hbar m}.
\end{aligned}
\end{equation}

\subsection{Single Particle Decay}\label{sec:singleDecay}

In the two-state approximation for momentum, the spontaneous emission breaks permutational symmetry.
To calculate the final jump operators in the two mode picture, we preemptively define single particle pseudo-spin operators;
\begin{equation}
\exp(\pm i 2 k \hat{z}_\mu) \approx \hat{s}^\pm_\mu \equiv \ket{+2\hbar k}_\mu \bra{0 \hbar k}_\mu,
\end{equation}

Furthermore, single particle spontaneous emission doesn't stay on the momentum grid like coherent dynamics or collective decay.
To treat this, we assume that
\begin{equation}
    \mathcal{N}(u) = A ( \delta(u+1) + \delta(u-1) ) + B \delta(u),
\end{equation}
for $u = \cos(\theta)$ and where $A$ and $B$ are chosen such that the first and second moments of the momentum-recoil matches the continuous case~\cite{Thesis_Bartolotta, Doppler_Castin}. 
With two free variables, $A$ and $B$, this is the best we can do.
This yields $A = 1/5$ and $B = 3/5$, which means the original spontaneous emission operator in Eq.~\eqref{eq:spontemiss} becomes
\begin{equation}
\begin{aligned}
\hat{\mathcal{D}}[\hat{L}_\mu] \hat{\rho} = & \frac{\gamma}{5} \left( \exp(-i k \hat{z}_\mu) \hat{\sigma}^-_\mu \hat{\rho} \hat{\sigma}^+_\mu \exp(+i k \hat{z}_\mu) \right) +... \\ 
& + \frac{\gamma}{5} \left( \exp(+i k \hat{z}_\mu) \hat{\sigma}^-_\mu \hat{\rho} \hat{\sigma}^+_\mu \exp(-i k \hat{z}_\mu) \right)+... \\
& + \frac{3 \gamma}{5} \left( \hat{\sigma}^-_\mu \hat{\rho} \hat{\sigma}^+_\mu \right) - \frac{\gamma}{2} \{ \hat{\sigma}^+_\mu \hat{\sigma}^-_\mu, \hat{\rho} \}\\
= & \hat{\mathcal{D}}[\hat{L}_\mu^{+1}] \hat{\rho} + \hat{\mathcal{D}}[\hat{L}_\mu^{-1}] \hat{\rho} + \hat{\mathcal{D}}[\hat{L}_\mu^{0}] \hat{\rho}.
\end{aligned}
\end{equation}
where 
\begin{equation}
\hat{L}_\mu^{\pm1} \equiv \sqrt{\frac{\gamma}{5}} \exp(\pm i k \hat{z}_\mu) \hat{\sigma}^-_\mu, \quad \mathrm{and} \quad 
\hat{L}_\mu^{0} \equiv \sqrt{\frac{3 \gamma}{5}} \hat{\sigma}^-_\mu .
\end{equation}

After eliminating the excited state we find
\begin{equation}
\tilde{\hat{L}}_{\mathrm{rec},\mu}(\theta) = \sqrt{2 \gamma U_0/\Delta_a } \ \hat{d}_\mu(\theta) ( \beta + \tilde{a}) \cos( k \hat{z}_\mu ),
\end{equation}
but we assume that spontaneous emission is only due to absorption from the coherent field, so that $( \beta + \tilde{a}) \cos( k \hat{z}_\mu ) \approx \beta \cos( k \hat{z}_\mu )$. 
This yields
\begin{equation}
\hat{L}_\mu \equiv \sqrt{ \frac{\gamma U_0 |\beta|^2 \mathcal{N}(u)}{(\Delta_a/2)} } \ \cos(k \hat{z}_\mu) \exp(- i k \hat{z}_\mu u )
\end{equation}
which, in the discrete momentum family, becomes
\begin{equation}
\begin{aligned}
\hat{L}_\mu^{\pm1} &\equiv \sqrt{ \frac{2 \gamma U_0 |\beta|^2 }{5 \Delta_a} } \ \cos(k \hat{z}_\mu) \exp(\pm i k \hat{z}_\mu ), \\
\hat{L}_\mu^{0} &\equiv  \sqrt{ \frac{6 |\beta|^2  \gamma U_0 }{5 \Delta_a} } \ \cos(k \hat{z}_\mu).
\end{aligned}
\end{equation}
Here, we note that $\hat{L}^0_\mu$ acting on the state $0 \hbar k$ or $+2 \hbar k$ always leads to atom loss from the 2-level manifold.
$\hat{L}^{\pm1}$, however, does not:
\begin{equation}
\begin{aligned}
\cos(k \hat{z}_\mu) \exp(\pm i k \hat{z}_\mu ) =& \frac{1}{2} \left( \exp(\pm i 2 k \hat{z}_\mu) + 1 \right) \\
=& \frac{1}{2} \left( \hat{s}_\mu^{\pm} + 1 \right) \\
\end{aligned}
\end{equation}
and, therefore,
\begin{equation} \label{eq:SingleEmission}
\hat{L}_\mu^{\pm1} \equiv  \sqrt{ \frac{\gamma U_0 |\beta|^2 }{(10 \Delta_a)} } \ ( \hat{s}^{\pm}_\mu + 1 ), \quad \hat{L}_\mu^{0} \equiv  \sqrt{ \frac{3 \gamma U_0 |\beta|^2 }{(10 \Delta_a)} } \ \hat{\ell}_\mu.
\end{equation}
where $\hat{\ell}_\mu \approx 2 \cos( k \hat{z}_\mu)$ represents atom loss into states like $\ket{\pm \frac{1}{2} 2 \hbar k}$.
The approximation that an atom in these states is lost is valid for short times because while these states remain relatively un-populated.

\section{Optimal Spin Squeezing Parameter}\label{sec:SpinSqueezing}
We take the one axis twisting Hamiltonian from the main text, $\hat{H}_\mathrm{OAT}$ and ignore the many-body energy gap, $\hat{J}^2$;
\begin{equation}
\begin{aligned}
\hat{H}_{\mathrm{eff}} =& \hbar \chi_0 \hat{J}_z^2, \quad \hat{L}_\pm \equiv \sqrt{\Gamma_\pm} \ \hat{J}^\pm, \quad \hat{L}_\mu \equiv \sqrt{\gamma_l} \hat{\ell}_j ,
\end{aligned}
\end{equation}
where
\begin{equation}
\begin{aligned}
\chi_0 \equiv& \frac{-\chi}{2}, \quad \Gamma_\pm \equiv \frac{\kappa |\beta|^2 |A^\pm|^2 }{4},\\
\mathrm{and}& \quad \gamma_l \equiv \frac{\gamma U_0 |\beta|^2}{2 \Delta_a},
\end{aligned}
\end{equation}
for particle loss rate $\gamma_l$ found in Eq.~\eqref{eq:SingleEmission}.
As an aside, we have assumed that all three decay channels lead to particle loss--which strictly over-estimates the total decay.

The squeezing parameter is given by:
\begin{equation}
\xi^2 = \min_\theta \frac{N (\Delta J^\perp_\theta)^2}{|\langle \textbf{J} \rangle|^2}.
\end{equation}
For short squeezing times the average pseudo-spin is in the $x$-direction, and the length goes as, $\langle \textbf{J} \rangle = S \mathrm{e}^{-\gamma_l t}$, for $S \equiv N/2$ being the total initial pseuo-spin length.
We note that $ |\langle \textbf{J} \rangle |^2 / N = S \mathrm{e}^{- 2 \gamma_l t}/2 $.

The minimum over $\theta$ is the smallest eigenvalue of the covariance matrix;
\begin{equation}
\textbf{C} = \begin{pmatrix}
\mathrm{var}(\hat{J}_y) & \mathrm{cov}(\hat{J}_y,\hat{J}_z) \\
\mathrm{cov}(\hat{J}_z,\hat{J}_y) & \mathrm{var}(\hat{J}_z)
\end{pmatrix}
\end{equation}
for $\mathrm{cov}(\hat{A}, \hat{B}) \equiv \exv{ \{ \hat{A} , \hat{B} \} }_{\psi}/2 - \exv{ \hat{A} }_{\psi} \exv{ \hat{B} }_{\psi}$ and $\mathrm{var}(\hat{A}) = \mathrm{cov}(\hat{A},\hat{A})$.
The state is initially polarized in $x$ and therefore $\langle \hat{J}_y \rangle = \langle \hat{J}_z \rangle = 0$ over short times.
The eigenvalues of $\textbf{C}$ are:
\begin{equation}
V_\pm = \frac{1}{2} \left( \mathcal{A}(t) + \mathcal{B}(t) \pm \sqrt{ \left( \mathcal{B}(t) - \mathcal{A}(t) \right)^2 +\mathcal{C}(t)^2} \right),
\end{equation}
where
\begin{equation}
\begin{aligned}
\mathcal{A}(t) &= \Delta J_y^2, \quad \mathcal{B}(t) =\Delta J_z^2, \\ 
& \mathcal{C}(t) = \langle \{ \hat{J}_y, \hat{J}_z \} \rangle, 
\end{aligned}
\end{equation}
For short times, the OAT dynamics can be treated independently of collective decay, and fluctuations from collective emission are added in quadrature to the fluctuations due to squeezing.
We first solve for the spin squeezing parameter in the collcetive case, $(\xi_0)^2$, and then add the effects of collective decay back in by modifying $\Delta J_z^2 $, with an additional term for each collective decay channel:
\begin{equation}
\begin{aligned}
T(\Gamma_\pm) &= S \tanh(2 S \Gamma_\pm t) ( 1 - \tanh(2 S \Gamma_\pm t) ) \\
\Delta J_z^2 &\rightarrow \Delta J_z^2 + T(\Gamma_+) + T(\Gamma_-),
\end{aligned}
\end{equation}
which requires that $N\Gamma_+ t$, $N \Gamma_- t \ll 1$.

Without collective decay, we find~\cite{Squeeze_Ueda}:
\begin{equation}
\begin{aligned}
\Delta J_y^2 =& \frac{S \mathrm{e}^{-\gamma_l t}}{2} \\
& + \frac{S}{2} \left(S- \frac{1}{2}\right) \left( 1 - \cos^{(2 S - 2)}(2 \chi_0 t) \right) \mathrm{e}^{-2 \gamma_l t} \\
\Delta J_z^2 =& \frac{S \mathrm{e}^{-\gamma_l t}}{2} \\
\langle \{ \hat{J}_y, \hat{J}_z \} \rangle =& 2 S \left(S- \frac{1}{2}\right) \sin(\chi_0 t) \cos^{(2 S - 1)}(\chi_0 t) \mathrm{e}^{-2 \gamma_l t}.
\end{aligned}
\end{equation}

For intermediate times, we have that $\mathcal{A} \gg \mathcal{B}, \mathcal{C}$, and therefore
\begin{equation}
\begin{aligned}
V_- =&  \frac{1}{2} \left( \mathcal{A}(t) + \mathcal{B}(t) - \sqrt{ \left( \mathcal{B}(t) - \mathcal{A}(t) \right)^2 + \mathcal{C}(t)^2} \right) \\
\approx& \frac{4 \mathcal{A} \mathcal{B} - \mathcal{C}^2}{4 \mathcal{A}} \\
\end{aligned}
\end{equation}
so
\begin{equation}
\xi^2 = N \frac{4 \mathcal{A} \mathcal{B} - \mathcal{C}^2}{4 \mathcal{A} {|\langle \textbf{J} \rangle|^2}}.
\end{equation}

Now, since we are considering short squeezing times we have small $\chi_0 t$, and approximate $\cos(\chi_0 t) \approx \mathrm{e}^{- (\chi_0 t)^2 / 2}$.
We also use large $N$ to drop terms $S+y=S$ for half or whole small integer $y$.
\begin{equation}
\begin{aligned}
& \Delta J_y^2 = \frac{S \mathrm{e}^{-\gamma_l t}}{2} + \frac{S^2}{2} \left( 1 - \mathrm{e}^{- 4 S (\chi_0 t)^2)} \right) \mathrm{e}^{-2 \gamma_l t} \\
& \Delta J_z^2 = \frac{S \mathrm{e}^{-\gamma_l t}}{2} \\
& \langle \{ \hat{J}_y, \hat{J}_z \} \rangle = 2 S^2 \chi_0 t \mathrm{e}^{-S (\chi_0 t)^2} \mathrm{e}^{-2 \gamma_l t}.
\end{aligned}
\end{equation}
We identify $q = S ( \chi_0 t)^2$ as a small unitless parameter and series expand to $\mathcal{O}(q^3)$.
\begin{equation}
\begin{aligned}
\Delta J_y^2 =&  \frac{S \mathrm{e}^{-\gamma_l t}}{2} + \frac{S^2}{2} \left( 1 - \mathrm{e}^{- 4 q} \right) \mathrm{e}^{-2 \gamma_l t} \\
\approx& \frac{S \mathrm{e}^{-\gamma_l t}}{2} + 2 S^2 q \left( 1 - 2 q + \frac{8}{3} q^2 \right) \mathrm{e}^{-2 \gamma_l t} \\ 
\langle \{ \hat{J}_y, \hat{J}_z \} \rangle^2 =& 4 S^3 q \mathrm{e}^{-2 q} \mathrm{e}^{-4 \gamma_l t} \\
\approx & 4 S^3 q \mathrm{e}^{-4 \gamma_l t} \left( 1 - 2 q + 2 q^2 \right).
\end{aligned}
\end{equation}
This all yields
\begin{equation}
\begin{aligned}
4 \mathcal{A} \mathcal{B} - \mathcal{C}^2 =& S^2 \mathrm{e}^{-2 \gamma_l t} + 4 S^3 q \left( 1 - 2 q + \frac{8}{3} q^2 \right) \mathrm{e}^{-3 \gamma_l t} \\
& - 4 S^3 q \mathrm{e}^{-4 \gamma_l t} \left( 1 - 2 q + 2 q^2 \right)
\end{aligned}
\end{equation}
and
\begin{equation}
\begin{aligned}
\mathcal{A} =&  \frac{S \mathrm{e}^{-\gamma_l t}}{2} + 2 S^2 q \left( 1 - 2 q + \frac{8}{3} q^2 \right) \mathrm{e}^{-2 \gamma_l t} \\
\approx& 2 S^2 q \mathrm{e}^{-2 \gamma_l t} .
\end{aligned}
\end{equation}
Now, we collect terms and drop them according to $S^2 q \gg S$ for strongly squeezed quadratures, yielding
\begin{equation}
\begin{aligned}
\xi^2 =& \frac{4 \mathcal{A} \mathcal{B} - \mathcal{C}^2}{4 \mathcal{A} {|\langle \textbf{J} \rangle|^2/N}} \\
\approx & \frac{\mathrm{e}^{+2 \gamma_l t}}{4 S q} + \left( \mathrm{e}^{+\gamma_l t} - 1 \right)
\end{aligned}
\end{equation}
where, because we aren't near the limit of oversqueezed states, we don't have the typical curvature term $\propto q^2$.
Lastly, we add back in the term for collective emission.
In the limit of large $N$, we have that $V_- \approx \mathcal{B} = \Delta J_z^2$, and therefore
\begin{equation}
\xi^2 \approx (\xi_0)^2 + \frac{N ( T(\Gamma_+) + T(\Gamma_-) ) }{|\langle \textbf{J} \rangle|^2}
\end{equation}
where $(\xi_0)^2$ is the spin squeezing parameter without collective decay.
This yields
\begin{equation}
\begin{aligned}
\xi^2 &\approx \frac{\mathrm{e}^{+2 \gamma_l t}}{4 S q} + \left( \mathrm{e}^{+\gamma_l t} - 1 \right) + \frac{N ( T(\Gamma_+) + T(\Gamma_-) ) }{|\langle \textbf{J} \rangle|^2} \\
&= \text{e}^{2 \gamma_l t} \left( \frac{1}{4 S q} + 2\frac{T(\Gamma_+) +T(\Gamma_-) }{S} \right) + \left( \mathrm{e}^{+\gamma_l t} - 1 \right).
\end{aligned}
\end{equation}
Where, lastly, we series expand to first order in $\Gamma_\pm, \gamma_l$.
\begin{equation}
\begin{aligned}
\xi^2 &\approx \frac{1}{4 S q} + t \gamma_l + 4 S t \left( \Gamma_+ + \Gamma_- \right),
\end{aligned}
\end{equation}
yielding, as a final expression,
\begin{equation}
\begin{aligned}
\xi^2 &= \frac{1}{(N \chi_0 t)^2} + t \gamma_l + 2 N t \Gamma_c,
\end{aligned}
\end{equation}
where $\Gamma_c = \Gamma_+ + \Gamma_-$ is the collective decay defined in the main text.

As a reminder:
\begin{equation}
\begin{aligned}
\gamma_l \equiv& \frac{\gamma U_0 |\beta|^2  }{2 \Delta_a}, \quad
\Gamma_\pm \equiv \frac{\kappa  |\beta|^2 |A^\pm|^2}{4}, \\
|\beta|^2 \equiv& \frac{\eta^2}{(\Delta_c')^2 + \frac{\kappa^2}{4} }, \quad A^{\pm} = \frac{U_0}{\Delta_c' \pm \omega_z - \frac{i \kappa}{2} }.
\end{aligned}
\end{equation}
When only collective decay is considered, 
\begin{equation}
 \frac{1}{(N \chi_0 t)^2}, 2 N t \Gamma_c \gg t \gamma_l
\end{equation}
and one finds~\cite{RobustSpinSqueeze_Rey}
\begin{equation}
\xi^2 \approx 3 \left(\frac{\Gamma_c}{2 \chi} \right)^{2/3}, \quad t_\mathrm{opt} = \frac{1}{ N (\chi  \Gamma_c)^{1/3} },
\end{equation}
which is independent of $N$.

One can improve this by considering longer timescales and solving for the dressed cavity detuning such that the coopertivity is of order 1;
\begin{equation}
\gamma_l = 2 N \Gamma_c
\end{equation}
so we find
\begin{equation}
\Delta'_{c,\mathrm{opt}} \approx \pm  \sqrt{ \frac{2 N U_0 \Delta_a \kappa}{\gamma} }.
\end{equation}
where the positive or negative solutions correspond to picking the peak from the red or blue shifted sidebands, $A^+$ or $A^-$, respectively.

This yields
\begin{equation}
\begin{aligned}
\xi^2 =& 2\gamma_l t + \frac{1}{(N \chi_0 t)^2 }
\end{aligned}
\end{equation}
which we optimize with respect to $t$ to find
\begin{equation}
\begin{aligned}
& t_\mathrm{opt} = \left( \gamma_l\ (N\chi_0)^{2} \right)^{-1/3},
\end{aligned}
\end{equation}
which, to highest order in $N$, is
\begin{equation}
\begin{aligned}
t_\mathrm{opt} \approx 4 \left( \frac{2 N^2 \Delta_a^5 \kappa^4}{ U_0 \gamma^5 \eta^6 } \right)^{1/3}
\end{aligned}
\end{equation}
giving
\begin{equation}
\begin{aligned}
\xi^2 \approx& 3 \left( \frac{2 \gamma \kappa }{N U_0 \Delta_a} \right)^{1/3} .
\end{aligned}
\end{equation}


\begin{thebibliography}{50}%
\makeatletter
\providecommand \@ifxundefined [1]{%
 \@ifx{#1\undefined}
}%
\providecommand \@ifnum [1]{%
 \ifnum #1\expandafter \@firstoftwo
 \else \expandafter \@secondoftwo
 \fi
}%
\providecommand \@ifx [1]{%
 \ifx #1\expandafter \@firstoftwo
 \else \expandafter \@secondoftwo
 \fi
}%
\providecommand \natexlab [1]{#1}%
\providecommand \enquote  [1]{``#1''}%
\providecommand \bibnamefont  [1]{#1}%
\providecommand \bibfnamefont [1]{#1}%
\providecommand \citenamefont [1]{#1}%
\providecommand \href@noop [0]{\@secondoftwo}%
\providecommand \href [0]{\begingroup \@sanitize@url \@href}%
\providecommand \@href[1]{\@@startlink{#1}\@@href}%
\providecommand \@@href[1]{\endgroup#1\@@endlink}%
\providecommand \@sanitize@url [0]{\catcode `\\12\catcode `\$12\catcode `\&12\catcode `\#12\catcode `\^12\catcode `\_12\catcode `\%12\relax}%
\providecommand \@@startlink[1]{}%
\providecommand \@@endlink[0]{}%
\providecommand \url  [0]{\begingroup\@sanitize@url \@url }%
\providecommand \@url [1]{\endgroup\@href {#1}{\urlprefix }}%
\providecommand \urlprefix  [0]{URL }%
\providecommand \Eprint [0]{\href }%
\providecommand \doibase [0]{https://doi.org/}%
\providecommand \selectlanguage [0]{\@gobble}%
\providecommand \bibinfo  [0]{\@secondoftwo}%
\providecommand \bibfield  [0]{\@secondoftwo}%
\providecommand \translation [1]{[#1]}%
\providecommand \BibitemOpen [0]{}%
\providecommand \bibitemStop [0]{}%
\providecommand \bibitemNoStop [0]{.\EOS\space}%
\providecommand \EOS [0]{\spacefactor3000\relax}%
\providecommand \BibitemShut  [1]{\csname bibitem#1\endcsname}%
\let\auto@bib@innerbib\@empty
%</preamble>
\bibitem [{\citenamefont {Luo}\ \emph {et~al.}(2024{\natexlab{a}})\citenamefont {Luo}, \citenamefont {Zhang}, \citenamefont {Koh}, \citenamefont {Wilson}, \citenamefont {Chu}, \citenamefont {Holland}, \citenamefont {Rey},\ and\ \citenamefont {Thompson}}]{MomentumExchange_Luo}%
  \BibitemOpen
  \bibfield  {author} {\bibinfo {author} {\bibfnamefont {C.}~\bibnamefont {Luo}}, \bibinfo {author} {\bibfnamefont {H.}~\bibnamefont {Zhang}}, \bibinfo {author} {\bibfnamefont {V.~P.}\ \bibnamefont {Koh}}, \bibinfo {author} {\bibfnamefont {J.~D.}\ \bibnamefont {Wilson}}, \bibinfo {author} {\bibfnamefont {A.}~\bibnamefont {Chu}}, \bibinfo {author} {\bibfnamefont {M.~J.}\ \bibnamefont {Holland}}, \bibinfo {author} {\bibfnamefont {A.~M.}\ \bibnamefont {Rey}},\ and\ \bibinfo {author} {\bibfnamefont {J.~K.}\ \bibnamefont {Thompson}},\ }\href@noop {} {\bibfield  {journal} {\bibinfo  {journal} {Science}\ }\textbf {\bibinfo {volume} {384}},\ \bibinfo {pages} {551} (\bibinfo {year} {2024}{\natexlab{a}})}\BibitemShut {NoStop}%
\bibitem [{\citenamefont {Ludlow}\ \emph {et~al.}(2015)\citenamefont {Ludlow}, \citenamefont {Boyd}, \citenamefont {Ye}, \citenamefont {Peik},\ and\ \citenamefont {Schmidt}}]{Clock_Ludlow}%
  \BibitemOpen
  \bibfield  {author} {\bibinfo {author} {\bibfnamefont {A.~D.}\ \bibnamefont {Ludlow}}, \bibinfo {author} {\bibfnamefont {M.~M.}\ \bibnamefont {Boyd}}, \bibinfo {author} {\bibfnamefont {J.}~\bibnamefont {Ye}}, \bibinfo {author} {\bibfnamefont {E.}~\bibnamefont {Peik}},\ and\ \bibinfo {author} {\bibfnamefont {P.~O.}\ \bibnamefont {Schmidt}},\ }\href {https://doi.org/10.1103/RevModPhys.87.637} {\bibfield  {journal} {\bibinfo  {journal} {Rev. Mod. Phys.}\ }\textbf {\bibinfo {volume} {87}},\ \bibinfo {pages} {637} (\bibinfo {year} {2015})}\BibitemShut {NoStop}%
\bibitem [{\citenamefont {Kessler}\ \emph {et~al.}(2014)\citenamefont {Kessler}, \citenamefont {K\'om\'ar}, \citenamefont {Bishof}, \citenamefont {Jiang}, \citenamefont {S\o{}rensen}, \citenamefont {Ye},\ and\ \citenamefont {Lukin}}]{Clock_Lukin}%
  \BibitemOpen
  \bibfield  {author} {\bibinfo {author} {\bibfnamefont {E.~M.}\ \bibnamefont {Kessler}}, \bibinfo {author} {\bibfnamefont {P.}~\bibnamefont {K\'om\'ar}}, \bibinfo {author} {\bibfnamefont {M.}~\bibnamefont {Bishof}}, \bibinfo {author} {\bibfnamefont {L.}~\bibnamefont {Jiang}}, \bibinfo {author} {\bibfnamefont {A.~S.}\ \bibnamefont {S\o{}rensen}}, \bibinfo {author} {\bibfnamefont {J.}~\bibnamefont {Ye}},\ and\ \bibinfo {author} {\bibfnamefont {M.~D.}\ \bibnamefont {Lukin}},\ }\href {https://doi.org/10.1103/PhysRevLett.112.190403} {\bibfield  {journal} {\bibinfo  {journal} {Phys. Rev. Lett.}\ }\textbf {\bibinfo {volume} {112}},\ \bibinfo {pages} {190403} (\bibinfo {year} {2014})}\BibitemShut {NoStop}%
\bibitem [{\citenamefont {Taylor}\ and\ \citenamefont {Bowen}(2016)}]{Bio_Taylor}%
  \BibitemOpen
  \bibfield  {author} {\bibinfo {author} {\bibfnamefont {M.~A.}\ \bibnamefont {Taylor}}\ and\ \bibinfo {author} {\bibfnamefont {W.~P.}\ \bibnamefont {Bowen}},\ }\href {https://doi.org/https://doi.org/10.1016/j.physrep.2015.12.002} {\bibfield  {journal} {\bibinfo  {journal} {Physics Reports}\ }\textbf {\bibinfo {volume} {615}},\ \bibinfo {pages} {1} (\bibinfo {year} {2016})},\ \bibinfo {note} {quantum metrology and its application in biology}\BibitemShut {NoStop}%
\bibitem [{\citenamefont {Qvarfort}\ \emph {et~al.}(2018)\citenamefont {Qvarfort}, \citenamefont {Serafini}, \citenamefont {Barker},\ and\ \citenamefont {Bose}}]{Gavimetry_Qvarfort}%
  \BibitemOpen
  \bibfield  {author} {\bibinfo {author} {\bibfnamefont {S.}~\bibnamefont {Qvarfort}}, \bibinfo {author} {\bibfnamefont {A.}~\bibnamefont {Serafini}}, \bibinfo {author} {\bibfnamefont {P.~F.}\ \bibnamefont {Barker}},\ and\ \bibinfo {author} {\bibfnamefont {S.}~\bibnamefont {Bose}},\ }\href {https://doi.org/10.1038/s41467-018-06037-z} {\bibfield  {journal} {\bibinfo  {journal} {Nature Communications}\ }\textbf {\bibinfo {volume} {9}},\ \bibinfo {pages} {3690} (\bibinfo {year} {2018})}\BibitemShut {NoStop}%
\bibitem [{\citenamefont {Richardson}\ \emph {et~al.}(2020)\citenamefont {Richardson}, \citenamefont {Hines}, \citenamefont {Schaffer}, \citenamefont {Anderson},\ and\ \citenamefont {Guzmán}}]{Optomechanical_Richardson}%
  \BibitemOpen
  \bibfield  {author} {\bibinfo {author} {\bibfnamefont {L.}~\bibnamefont {Richardson}}, \bibinfo {author} {\bibfnamefont {A.}~\bibnamefont {Hines}}, \bibinfo {author} {\bibfnamefont {A.}~\bibnamefont {Schaffer}}, \bibinfo {author} {\bibfnamefont {B.}~\bibnamefont {Anderson}},\ and\ \bibinfo {author} {\bibfnamefont {F.}~\bibnamefont {Guzmán}},\ }\href {https://doi.org/10.1364/AO.393060} {\bibfield  {journal} {\bibinfo  {journal} {Applied Optics}\ }\textbf {\bibinfo {volume} {59}} (\bibinfo {year} {2020})}\BibitemShut {NoStop}%
\bibitem [{\citenamefont {Templier}\ \emph {et~al.}(2022)\citenamefont {Templier}, \citenamefont {Cheiney}, \citenamefont {d’Armagnac~de Castanet}, \citenamefont {Gouraud}, \citenamefont {Porte}, \citenamefont {Napolitano}, \citenamefont {Bouyer}, \citenamefont {Battelier},\ and\ \citenamefont {Barrett}}]{Accelerometer_Templier}%
  \BibitemOpen
  \bibfield  {author} {\bibinfo {author} {\bibfnamefont {S.}~\bibnamefont {Templier}}, \bibinfo {author} {\bibfnamefont {P.}~\bibnamefont {Cheiney}}, \bibinfo {author} {\bibfnamefont {Q.}~\bibnamefont {d’Armagnac~de Castanet}}, \bibinfo {author} {\bibfnamefont {B.}~\bibnamefont {Gouraud}}, \bibinfo {author} {\bibfnamefont {H.}~\bibnamefont {Porte}}, \bibinfo {author} {\bibfnamefont {F.}~\bibnamefont {Napolitano}}, \bibinfo {author} {\bibfnamefont {P.}~\bibnamefont {Bouyer}}, \bibinfo {author} {\bibfnamefont {B.}~\bibnamefont {Battelier}},\ and\ \bibinfo {author} {\bibfnamefont {B.}~\bibnamefont {Barrett}},\ }\href {https://doi.org/10.1126/sciadv.add3854} {\bibfield  {journal} {\bibinfo  {journal} {Science Advances}\ }\textbf {\bibinfo {volume} {8}},\ \bibinfo {pages} {eadd3854} (\bibinfo {year} {2022})}\BibitemShut {NoStop}%
\bibitem [{\citenamefont {Bothwell}\ \emph {et~al.}(2022)\citenamefont {Bothwell}, \citenamefont {Kennedy}, \citenamefont {Aeppli}, \citenamefont {Kedar}, \citenamefont {Robinson}, \citenamefont {Oelker}, \citenamefont {Staron},\ and\ \citenamefont {Ye}}]{RedShift_Bothwell}%
  \BibitemOpen
  \bibfield  {author} {\bibinfo {author} {\bibfnamefont {T.}~\bibnamefont {Bothwell}}, \bibinfo {author} {\bibfnamefont {C.~J.}\ \bibnamefont {Kennedy}}, \bibinfo {author} {\bibfnamefont {A.}~\bibnamefont {Aeppli}}, \bibinfo {author} {\bibfnamefont {D.}~\bibnamefont {Kedar}}, \bibinfo {author} {\bibfnamefont {J.~M.}\ \bibnamefont {Robinson}}, \bibinfo {author} {\bibfnamefont {E.}~\bibnamefont {Oelker}}, \bibinfo {author} {\bibfnamefont {A.}~\bibnamefont {Staron}},\ and\ \bibinfo {author} {\bibfnamefont {J.}~\bibnamefont {Ye}},\ }\href@noop {} {\bibfield  {journal} {\bibinfo  {journal} {Nature}\ }\textbf {\bibinfo {volume} {602}},\ \bibinfo {pages} {420} (\bibinfo {year} {2022})}\BibitemShut {NoStop}%
\bibitem [{\citenamefont {Aasi}\ \emph {et~al.}(2013)\citenamefont {Aasi}, \citenamefont {Abadie}, \citenamefont {Abbott}, \citenamefont {Abbott}, \citenamefont {Abbott}, \citenamefont {Abernathy}, \citenamefont {Adams}, \citenamefont {Adams}, \citenamefont {Addesso}, \citenamefont {Adhikari} \emph {et~al.}}]{LIGO_Aasi}%
  \BibitemOpen
  \bibfield  {author} {\bibinfo {author} {\bibfnamefont {J.}~\bibnamefont {Aasi}}, \bibinfo {author} {\bibfnamefont {J.}~\bibnamefont {Abadie}}, \bibinfo {author} {\bibfnamefont {B.}~\bibnamefont {Abbott}}, \bibinfo {author} {\bibfnamefont {R.}~\bibnamefont {Abbott}}, \bibinfo {author} {\bibfnamefont {T.}~\bibnamefont {Abbott}}, \bibinfo {author} {\bibfnamefont {M.}~\bibnamefont {Abernathy}}, \bibinfo {author} {\bibfnamefont {C.}~\bibnamefont {Adams}}, \bibinfo {author} {\bibfnamefont {T.}~\bibnamefont {Adams}}, \bibinfo {author} {\bibfnamefont {P.}~\bibnamefont {Addesso}}, \bibinfo {author} {\bibfnamefont {R.}~\bibnamefont {Adhikari}}, \emph {et~al.},\ }\href@noop {} {\bibfield  {journal} {\bibinfo  {journal} {Nature Photonics}\ }\textbf {\bibinfo {volume} {7}},\ \bibinfo {pages} {613} (\bibinfo {year} {2013})}\BibitemShut {NoStop}%
\bibitem [{\citenamefont {Tse}\ \emph {et~al.}(2019)\citenamefont {Tse} \emph {et~al.}}]{LIGO_Tse}%
  \BibitemOpen
  \bibfield  {author} {\bibinfo {author} {\bibfnamefont {M.}~\bibnamefont {Tse}} \emph {et~al.},\ }\href {https://doi.org/10.1103/PhysRevLett.123.231107} {\bibfield  {journal} {\bibinfo  {journal} {Phys. Rev. Lett.}\ }\textbf {\bibinfo {volume} {123}},\ \bibinfo {pages} {231107} (\bibinfo {year} {2019})}\BibitemShut {NoStop}%
\bibitem [{\citenamefont {Abbott}\ \emph {et~al.}(2016{\natexlab{a}})\citenamefont {Abbott} \emph {et~al.}}]{BlackHoles_Abbott}%
  \BibitemOpen
  \bibfield  {author} {\bibinfo {author} {\bibfnamefont {B.~P.}\ \bibnamefont {Abbott}} \emph {et~al.} (\bibinfo {collaboration} {LIGO Scientific Collaboration and Virgo Collaboration}),\ }\href {https://doi.org/10.1103/PhysRevLett.116.061102} {\bibfield  {journal} {\bibinfo  {journal} {Phys. Rev. Lett.}\ }\textbf {\bibinfo {volume} {116}},\ \bibinfo {pages} {061102} (\bibinfo {year} {2016}{\natexlab{a}})}\BibitemShut {NoStop}%
\bibitem [{\citenamefont {Abbott}\ \emph {et~al.}(2016{\natexlab{b}})\citenamefont {Abbott} \emph {et~al.}}]{AdvancedBlackHoles_Abbott}%
  \BibitemOpen
  \bibfield  {author} {\bibinfo {author} {\bibfnamefont {B.~P.}\ \bibnamefont {Abbott}} \emph {et~al.} (\bibinfo {collaboration} {LIGO Scientific Collaboration and Virgo Collaboration}),\ }\href {https://doi.org/10.1103/PhysRevX.6.041015} {\bibfield  {journal} {\bibinfo  {journal} {Phys. Rev. X}\ }\textbf {\bibinfo {volume} {6}},\ \bibinfo {pages} {041015} (\bibinfo {year} {2016}{\natexlab{b}})}\BibitemShut {NoStop}%
\bibitem [{\citenamefont {Mandel}\ and\ \citenamefont {Wolf}(1995)}]{OpticalCoherenceText_mandel}%
  \BibitemOpen
  \bibfield  {author} {\bibinfo {author} {\bibfnamefont {L.}~\bibnamefont {Mandel}}\ and\ \bibinfo {author} {\bibfnamefont {E.}~\bibnamefont {Wolf}},\ }\href@noop {} {\emph {\bibinfo {title} {Optical coherence and quantum optics}}}\ (\bibinfo  {publisher} {Cambridge university press},\ \bibinfo {year} {1995})\BibitemShut {NoStop}%
\bibitem [{\citenamefont {Holland}\ and\ \citenamefont {Burnett}(1993)}]{HL_Holland}%
  \BibitemOpen
  \bibfield  {author} {\bibinfo {author} {\bibfnamefont {M.~J.}\ \bibnamefont {Holland}}\ and\ \bibinfo {author} {\bibfnamefont {K.}~\bibnamefont {Burnett}},\ }\href {https://doi.org/10.1103/PhysRevLett.71.1355} {\bibfield  {journal} {\bibinfo  {journal} {Phys. Rev. Lett.}\ }\textbf {\bibinfo {volume} {71}},\ \bibinfo {pages} {1355} (\bibinfo {year} {1993})}\BibitemShut {NoStop}%
\bibitem [{\citenamefont {Ma}\ \emph {et~al.}(2011)\citenamefont {Ma}, \citenamefont {Wang}, \citenamefont {Sun},\ and\ \citenamefont {Nori}}]{Squeezing_Ma}%
  \BibitemOpen
  \bibfield  {author} {\bibinfo {author} {\bibfnamefont {J.}~\bibnamefont {Ma}}, \bibinfo {author} {\bibfnamefont {X.}~\bibnamefont {Wang}}, \bibinfo {author} {\bibfnamefont {C.}~\bibnamefont {Sun}},\ and\ \bibinfo {author} {\bibfnamefont {F.}~\bibnamefont {Nori}},\ }\href {https://doi.org/https://doi.org/10.1016/j.physrep.2011.08.003} {\bibfield  {journal} {\bibinfo  {journal} {Physics Reports}\ }\textbf {\bibinfo {volume} {509}},\ \bibinfo {pages} {89} (\bibinfo {year} {2011})}\BibitemShut {NoStop}%
\bibitem [{\citenamefont {Kitagawa}\ and\ \citenamefont {Ueda}(1993)}]{Squeeze_Ueda}%
  \BibitemOpen
  \bibfield  {author} {\bibinfo {author} {\bibfnamefont {M.}~\bibnamefont {Kitagawa}}\ and\ \bibinfo {author} {\bibfnamefont {M.}~\bibnamefont {Ueda}},\ }\href {https://doi.org/10.1103/PhysRevA.47.5138} {\bibfield  {journal} {\bibinfo  {journal} {Phys. Rev. A}\ }\textbf {\bibinfo {volume} {47}},\ \bibinfo {pages} {5138} (\bibinfo {year} {1993})}\BibitemShut {NoStop}%
\bibitem [{\citenamefont {Pedrozo-Peñafiel}\ \emph {et~al.}(2020{\natexlab{a}})\citenamefont {Pedrozo-Peñafiel}, \citenamefont {Colombo}, \citenamefont {Shu}, \citenamefont {Adiyatullin}, \citenamefont {Li}, \citenamefont {Mendez}, \citenamefont {Braverman}, \citenamefont {Kawasaki}, \citenamefont {Akamatsu}, \citenamefont {Xiao},\ and\ \citenamefont {Vuleti\`c}}]{EntangledClock_Shu}%
  \BibitemOpen
  \bibfield  {author} {\bibinfo {author} {\bibfnamefont {E.}~\bibnamefont {Pedrozo-Peñafiel}}, \bibinfo {author} {\bibfnamefont {S.}~\bibnamefont {Colombo}}, \bibinfo {author} {\bibfnamefont {C.}~\bibnamefont {Shu}}, \bibinfo {author} {\bibfnamefont {A.~F.}\ \bibnamefont {Adiyatullin}}, \bibinfo {author} {\bibfnamefont {Z.}~\bibnamefont {Li}}, \bibinfo {author} {\bibfnamefont {E.}~\bibnamefont {Mendez}}, \bibinfo {author} {\bibfnamefont {B.}~\bibnamefont {Braverman}}, \bibinfo {author} {\bibfnamefont {A.}~\bibnamefont {Kawasaki}}, \bibinfo {author} {\bibfnamefont {D.}~\bibnamefont {Akamatsu}}, \bibinfo {author} {\bibfnamefont {Y.}~\bibnamefont {Xiao}},\ and\ \bibinfo {author} {\bibfnamefont {V.}~\bibnamefont {Vuleti\`c}},\ }\href {https://doi.org/10.1038/s41586-020-3006-1} {\bibfield  {journal} {\bibinfo  {journal} {Nature}\ }\textbf {\bibinfo {volume} {588}},\ \bibinfo {pages} {414–418} (\bibinfo {year} {2020}{\natexlab{a}})}\BibitemShut {NoStop}%
\bibitem [{\citenamefont {Kuzmich}\ \emph {et~al.}(2000)\citenamefont {Kuzmich}, \citenamefont {Mandel},\ and\ \citenamefont {Bigelow}}]{SpinSqueezeQND_Kuzmich}%
  \BibitemOpen
  \bibfield  {author} {\bibinfo {author} {\bibfnamefont {A.}~\bibnamefont {Kuzmich}}, \bibinfo {author} {\bibfnamefont {L.}~\bibnamefont {Mandel}},\ and\ \bibinfo {author} {\bibfnamefont {N.~P.}\ \bibnamefont {Bigelow}},\ }\href {https://doi.org/10.1103/PhysRevLett.85.1594} {\bibfield  {journal} {\bibinfo  {journal} {Phys. Rev. Lett.}\ }\textbf {\bibinfo {volume} {85}},\ \bibinfo {pages} {1594} (\bibinfo {year} {2000})}\BibitemShut {NoStop}%
\bibitem [{\citenamefont {Greve}\ \emph {et~al.}(2022)\citenamefont {Greve}, \citenamefont {Luo}, \citenamefont {Wu},\ and\ \citenamefont {Thompson}}]{EntangledMomentum_Greve}%
  \BibitemOpen
  \bibfield  {author} {\bibinfo {author} {\bibfnamefont {G.~P.}\ \bibnamefont {Greve}}, \bibinfo {author} {\bibfnamefont {C.}~\bibnamefont {Luo}}, \bibinfo {author} {\bibfnamefont {B.}~\bibnamefont {Wu}},\ and\ \bibinfo {author} {\bibfnamefont {J.~K.}\ \bibnamefont {Thompson}},\ }\bibfield  {journal} {\bibinfo  {journal} {Nature}\ }\textbf {\bibinfo {volume} {610}},\ \href {https://doi.org/10.1038/s41586-022-05197-9} {10.1038/s41586-022-05197-9} (\bibinfo {year} {2022})\BibitemShut {NoStop}%
\bibitem [{\citenamefont {Malia}\ \emph {et~al.}(2022)\citenamefont {Malia}, \citenamefont {Wu}, \citenamefont {Mart{\'i}nez-Rinc{\'o}n},\ and\ \citenamefont {Kasevich}}]{EntangledModes_Malia}%
  \BibitemOpen
  \bibfield  {author} {\bibinfo {author} {\bibfnamefont {B.~K.}\ \bibnamefont {Malia}}, \bibinfo {author} {\bibfnamefont {Y.}~\bibnamefont {Wu}}, \bibinfo {author} {\bibfnamefont {J.}~\bibnamefont {Mart{\'i}nez-Rinc{\'o}n}},\ and\ \bibinfo {author} {\bibfnamefont {M.~A.}\ \bibnamefont {Kasevich}},\ }\href {https://doi.org/10.1038/s41586-022-05363-z} {\bibfield  {journal} {\bibinfo  {journal} {Nature}\ }\textbf {\bibinfo {volume} {612}},\ \bibinfo {pages} {661} (\bibinfo {year} {2022})}\BibitemShut {NoStop}%
\bibitem [{\citenamefont {Georgescu}\ \emph {et~al.}(2014)\citenamefont {Georgescu}, \citenamefont {Ashhab},\ and\ \citenamefont {Nori}}]{QSim_Georgescu}%
  \BibitemOpen
  \bibfield  {author} {\bibinfo {author} {\bibfnamefont {I.~M.}\ \bibnamefont {Georgescu}}, \bibinfo {author} {\bibfnamefont {S.}~\bibnamefont {Ashhab}},\ and\ \bibinfo {author} {\bibfnamefont {F.}~\bibnamefont {Nori}},\ }\href@noop {} {\bibfield  {journal} {\bibinfo  {journal} {Reviews of Modern Physics}\ }\textbf {\bibinfo {volume} {86}},\ \bibinfo {pages} {153} (\bibinfo {year} {2014})}\BibitemShut {NoStop}%
\bibitem [{\citenamefont {Mei}\ \emph {et~al.}(2022)\citenamefont {Mei}, \citenamefont {Li}, \citenamefont {Wu}, \citenamefont {Cai}, \citenamefont {Wang}, \citenamefont {Yao}, \citenamefont {Zhou},\ and\ \citenamefont {Duan}}]{RabiHubbardSim_Mei}%
  \BibitemOpen
  \bibfield  {author} {\bibinfo {author} {\bibfnamefont {Q.-X.}\ \bibnamefont {Mei}}, \bibinfo {author} {\bibfnamefont {B.-W.}\ \bibnamefont {Li}}, \bibinfo {author} {\bibfnamefont {Y.-K.}\ \bibnamefont {Wu}}, \bibinfo {author} {\bibfnamefont {M.-L.}\ \bibnamefont {Cai}}, \bibinfo {author} {\bibfnamefont {Y.}~\bibnamefont {Wang}}, \bibinfo {author} {\bibfnamefont {L.}~\bibnamefont {Yao}}, \bibinfo {author} {\bibfnamefont {Z.-C.}\ \bibnamefont {Zhou}},\ and\ \bibinfo {author} {\bibfnamefont {L.-M.}\ \bibnamefont {Duan}},\ }\href {https://doi.org/10.1103/PhysRevLett.128.160504} {\bibfield  {journal} {\bibinfo  {journal} {Phys. Rev. Lett.}\ }\textbf {\bibinfo {volume} {128}},\ \bibinfo {pages} {160504} (\bibinfo {year} {2022})}\BibitemShut {NoStop}%
\bibitem [{\citenamefont {Steinert}\ \emph {et~al.}(2023)\citenamefont {Steinert}, \citenamefont {Osterholz}, \citenamefont {Eberhard}, \citenamefont {Festa}, \citenamefont {Lorenz}, \citenamefont {Chen}, \citenamefont {Trautmann},\ and\ \citenamefont {Gross}}]{NuetralAtomSim_Steinert}%
  \BibitemOpen
  \bibfield  {author} {\bibinfo {author} {\bibfnamefont {L.-M.}\ \bibnamefont {Steinert}}, \bibinfo {author} {\bibfnamefont {P.}~\bibnamefont {Osterholz}}, \bibinfo {author} {\bibfnamefont {R.}~\bibnamefont {Eberhard}}, \bibinfo {author} {\bibfnamefont {L.}~\bibnamefont {Festa}}, \bibinfo {author} {\bibfnamefont {N.}~\bibnamefont {Lorenz}}, \bibinfo {author} {\bibfnamefont {Z.}~\bibnamefont {Chen}}, \bibinfo {author} {\bibfnamefont {A.}~\bibnamefont {Trautmann}},\ and\ \bibinfo {author} {\bibfnamefont {C.}~\bibnamefont {Gross}},\ }\href {https://doi.org/10.1103/PhysRevLett.130.243001} {\bibfield  {journal} {\bibinfo  {journal} {Phys. Rev. Lett.}\ }\textbf {\bibinfo {volume} {130}},\ \bibinfo {pages} {243001} (\bibinfo {year} {2023})}\BibitemShut {NoStop}%
\bibitem [{\citenamefont {Cao}\ \emph {et~al.}(2023)\citenamefont {Cao}, \citenamefont {Li}, \citenamefont {Zhao}, \citenamefont {Guo}, \citenamefont {Qi}, \citenamefont {Chang}, \citenamefont {Zhou}, \citenamefont {Xu},\ and\ \citenamefont {Duan}}]{IonSim_Cao}%
  \BibitemOpen
  \bibfield  {author} {\bibinfo {author} {\bibfnamefont {M.-M.}\ \bibnamefont {Cao}}, \bibinfo {author} {\bibfnamefont {K.}~\bibnamefont {Li}}, \bibinfo {author} {\bibfnamefont {W.-D.}\ \bibnamefont {Zhao}}, \bibinfo {author} {\bibfnamefont {W.-X.}\ \bibnamefont {Guo}}, \bibinfo {author} {\bibfnamefont {B.-X.}\ \bibnamefont {Qi}}, \bibinfo {author} {\bibfnamefont {X.-Y.}\ \bibnamefont {Chang}}, \bibinfo {author} {\bibfnamefont {Z.-C.}\ \bibnamefont {Zhou}}, \bibinfo {author} {\bibfnamefont {Y.}~\bibnamefont {Xu}},\ and\ \bibinfo {author} {\bibfnamefont {L.-M.}\ \bibnamefont {Duan}},\ }\href {https://doi.org/10.1103/PhysRevLett.130.163001} {\bibfield  {journal} {\bibinfo  {journal} {Phys. Rev. Lett.}\ }\textbf {\bibinfo {volume} {130}},\ \bibinfo {pages} {163001} (\bibinfo {year} {2023})}\BibitemShut {NoStop}%
\bibitem [{\citenamefont {Leroux}\ \emph {et~al.}(2010)\citenamefont {Leroux}, \citenamefont {Schleier-Smith},\ and\ \citenamefont {Vuleti\ifmmode~\acute{c}\else \'{c}\fi{}}}]{Leroux}%
  \BibitemOpen
  \bibfield  {author} {\bibinfo {author} {\bibfnamefont {I.~D.}\ \bibnamefont {Leroux}}, \bibinfo {author} {\bibfnamefont {M.~H.}\ \bibnamefont {Schleier-Smith}},\ and\ \bibinfo {author} {\bibfnamefont {V.}~\bibnamefont {Vuleti\ifmmode~\acute{c}\else \'{c}\fi{}}},\ }\href {https://doi.org/10.1103/PhysRevLett.104.073602} {\bibfield  {journal} {\bibinfo  {journal} {Phys. Rev. Lett.}\ }\textbf {\bibinfo {volume} {104}},\ \bibinfo {pages} {073602} (\bibinfo {year} {2010})}\BibitemShut {NoStop}%
\bibitem [{\citenamefont {Pedrozo-Peñafiel}\ \emph {et~al.}(2020{\natexlab{b}})\citenamefont {Pedrozo-Peñafiel}, \citenamefont {Colombo}, \citenamefont {Shu}, \citenamefont {Adiyatullin}, \citenamefont {Li}, \citenamefont {Mendez}, \citenamefont {Braverman}, \citenamefont {Kawasaki}, \citenamefont {Akamatsu}, \citenamefont {Xiao},\ and\ \citenamefont {Vuleti\`c}}]{PedrozoPenafiel}%
  \BibitemOpen
  \bibfield  {author} {\bibinfo {author} {\bibfnamefont {E.}~\bibnamefont {Pedrozo-Peñafiel}}, \bibinfo {author} {\bibfnamefont {S.}~\bibnamefont {Colombo}}, \bibinfo {author} {\bibfnamefont {C.}~\bibnamefont {Shu}}, \bibinfo {author} {\bibfnamefont {A.~F.}\ \bibnamefont {Adiyatullin}}, \bibinfo {author} {\bibfnamefont {Z.}~\bibnamefont {Li}}, \bibinfo {author} {\bibfnamefont {E.}~\bibnamefont {Mendez}}, \bibinfo {author} {\bibfnamefont {B.}~\bibnamefont {Braverman}}, \bibinfo {author} {\bibfnamefont {A.}~\bibnamefont {Kawasaki}}, \bibinfo {author} {\bibfnamefont {D.}~\bibnamefont {Akamatsu}}, \bibinfo {author} {\bibfnamefont {Y.}~\bibnamefont {Xiao}},\ and\ \bibinfo {author} {\bibfnamefont {V.}~\bibnamefont {Vuleti\`c}},\ }\href {https://doi.org/10.1038/s41586-020-3006-1} {\bibfield  {journal} {\bibinfo  {journal} {Nature}\ }\textbf {\bibinfo {volume} {588}},\ \bibinfo {pages} {414–418} (\bibinfo {year} {2020}{\natexlab{b}})}\BibitemShut {NoStop}%
\bibitem [{\citenamefont {Reilly}\ \emph {et~al.}(2022)\citenamefont {Reilly}, \citenamefont {J\"ager}, \citenamefont {Cooper},\ and\ \citenamefont {Holland}}]{DFS_Reilly}%
  \BibitemOpen
  \bibfield  {author} {\bibinfo {author} {\bibfnamefont {J.~T.}\ \bibnamefont {Reilly}}, \bibinfo {author} {\bibfnamefont {S.~B.}\ \bibnamefont {J\"ager}}, \bibinfo {author} {\bibfnamefont {J.}~\bibnamefont {Cooper}},\ and\ \bibinfo {author} {\bibfnamefont {M.~J.}\ \bibnamefont {Holland}},\ }\href {https://doi.org/10.1103/PhysRevA.106.023703} {\bibfield  {journal} {\bibinfo  {journal} {Phys. Rev. A}\ }\textbf {\bibinfo {volume} {106}},\ \bibinfo {pages} {023703} (\bibinfo {year} {2022})}\BibitemShut {NoStop}%
\bibitem [{\citenamefont {Wilson}\ \emph {et~al.}(2022)\citenamefont {Wilson}, \citenamefont {J\"ager}, \citenamefont {Reilly}, \citenamefont {Shankar}, \citenamefont {Chiofalo},\ and\ \citenamefont {Holland}}]{SU4_Wilson}%
  \BibitemOpen
  \bibfield  {author} {\bibinfo {author} {\bibfnamefont {J.~D.}\ \bibnamefont {Wilson}}, \bibinfo {author} {\bibfnamefont {S.~B.}\ \bibnamefont {J\"ager}}, \bibinfo {author} {\bibfnamefont {J.~T.}\ \bibnamefont {Reilly}}, \bibinfo {author} {\bibfnamefont {A.}~\bibnamefont {Shankar}}, \bibinfo {author} {\bibfnamefont {M.~L.}\ \bibnamefont {Chiofalo}},\ and\ \bibinfo {author} {\bibfnamefont {M.~J.}\ \bibnamefont {Holland}},\ }\href {https://doi.org/10.1103/PhysRevA.106.043711} {\bibfield  {journal} {\bibinfo  {journal} {Phys. Rev. A}\ }\textbf {\bibinfo {volume} {106}},\ \bibinfo {pages} {043711} (\bibinfo {year} {2022})}\BibitemShut {NoStop}%
\bibitem [{\citenamefont {Reilly}\ \emph {et~al.}(2023)\citenamefont {Reilly}, \citenamefont {Wilson}, \citenamefont {J\"ager}, \citenamefont {Wilson},\ and\ \citenamefont {Holland}}]{OptimalGenerators_Wilson}%
  \BibitemOpen
  \bibfield  {author} {\bibinfo {author} {\bibfnamefont {J.~T.}\ \bibnamefont {Reilly}}, \bibinfo {author} {\bibfnamefont {J.~D.}\ \bibnamefont {Wilson}}, \bibinfo {author} {\bibfnamefont {S.~B.}\ \bibnamefont {J\"ager}}, \bibinfo {author} {\bibfnamefont {C.}~\bibnamefont {Wilson}},\ and\ \bibinfo {author} {\bibfnamefont {M.~J.}\ \bibnamefont {Holland}},\ }\href {https://doi.org/10.1103/PhysRevLett.131.150802} {\bibfield  {journal} {\bibinfo  {journal} {Phys. Rev. Lett.}\ }\textbf {\bibinfo {volume} {131}},\ \bibinfo {pages} {150802} (\bibinfo {year} {2023})}\BibitemShut {NoStop}%
\bibitem [{\citenamefont {Shankar}\ \emph {et~al.}(2019{\natexlab{a}})\citenamefont {Shankar}, \citenamefont {Salvi}, \citenamefont {Chiofalo}, \citenamefont {Poli},\ and\ \citenamefont {Holland}}]{Bragg_Shankar}%
  \BibitemOpen
  \bibfield  {author} {\bibinfo {author} {\bibfnamefont {A.}~\bibnamefont {Shankar}}, \bibinfo {author} {\bibfnamefont {L.}~\bibnamefont {Salvi}}, \bibinfo {author} {\bibfnamefont {M.~L.}\ \bibnamefont {Chiofalo}}, \bibinfo {author} {\bibfnamefont {N.}~\bibnamefont {Poli}},\ and\ \bibinfo {author} {\bibfnamefont {M.~J.}\ \bibnamefont {Holland}},\ }\href {https://doi.org/10.1088/2058-9565/ab455d} {\bibfield  {journal} {\bibinfo  {journal} {Quantum Science and Technology}\ }\textbf {\bibinfo {volume} {4}},\ \bibinfo {pages} {045010} (\bibinfo {year} {2019}{\natexlab{a}})}\BibitemShut {NoStop}%
\bibitem [{\citenamefont {Salvi}\ \emph {et~al.}(2018)\citenamefont {Salvi}, \citenamefont {Poli}, \citenamefont {Vuleti\ifmmode~\acute{c}\else \'{c}\fi{}},\ and\ \citenamefont {Tino}}]{MomentumSqueezing_Guglielmo}%
  \BibitemOpen
  \bibfield  {author} {\bibinfo {author} {\bibfnamefont {L.}~\bibnamefont {Salvi}}, \bibinfo {author} {\bibfnamefont {N.}~\bibnamefont {Poli}}, \bibinfo {author} {\bibfnamefont {V.}~\bibnamefont {Vuleti\ifmmode~\acute{c}\else \'{c}\fi{}}},\ and\ \bibinfo {author} {\bibfnamefont {G.~M.}\ \bibnamefont {Tino}},\ }\href {https://doi.org/10.1103/PhysRevLett.120.033601} {\bibfield  {journal} {\bibinfo  {journal} {Phys. Rev. Lett.}\ }\textbf {\bibinfo {volume} {120}},\ \bibinfo {pages} {033601} (\bibinfo {year} {2018})}\BibitemShut {NoStop}%
\bibitem [{\citenamefont {Mabuchi}\ \emph {et~al.}(1999)\citenamefont {Mabuchi}, \citenamefont {Ye},\ and\ \citenamefont {Kimble}}]{TCQED1999_Ye}%
  \BibitemOpen
  \bibfield  {author} {\bibinfo {author} {\bibfnamefont {H.}~\bibnamefont {Mabuchi}}, \bibinfo {author} {\bibfnamefont {J.}~\bibnamefont {Ye}},\ and\ \bibinfo {author} {\bibfnamefont {H.~J.}\ \bibnamefont {Kimble}},\ }\href@noop {} {\bibfield  {journal} {\bibinfo  {journal} {Applied Physics B}\ }\textbf {\bibinfo {volume} {68}},\ \bibinfo {pages} {1095} (\bibinfo {year} {1999})}\BibitemShut {NoStop}%
\bibitem [{\citenamefont {Fink}\ \emph {et~al.}(2009)\citenamefont {Fink}, \citenamefont {Bianchetti}, \citenamefont {Baur}, \citenamefont {G\"oppl}, \citenamefont {Steffen}, \citenamefont {Filipp}, \citenamefont {Leek}, \citenamefont {Blais},\ and\ \citenamefont {Wallraff}}]{TCQubits_Fink}%
  \BibitemOpen
  \bibfield  {author} {\bibinfo {author} {\bibfnamefont {J.~M.}\ \bibnamefont {Fink}}, \bibinfo {author} {\bibfnamefont {R.}~\bibnamefont {Bianchetti}}, \bibinfo {author} {\bibfnamefont {M.}~\bibnamefont {Baur}}, \bibinfo {author} {\bibfnamefont {M.}~\bibnamefont {G\"oppl}}, \bibinfo {author} {\bibfnamefont {L.}~\bibnamefont {Steffen}}, \bibinfo {author} {\bibfnamefont {S.}~\bibnamefont {Filipp}}, \bibinfo {author} {\bibfnamefont {P.~J.}\ \bibnamefont {Leek}}, \bibinfo {author} {\bibfnamefont {A.}~\bibnamefont {Blais}},\ and\ \bibinfo {author} {\bibfnamefont {A.}~\bibnamefont {Wallraff}},\ }\href {https://doi.org/10.1103/PhysRevLett.103.083601} {\bibfield  {journal} {\bibinfo  {journal} {Phys. Rev. Lett.}\ }\textbf {\bibinfo {volume} {103}},\ \bibinfo {pages} {083601} (\bibinfo {year} {2009})}\BibitemShut {NoStop}%
\bibitem [{\citenamefont {Norcia}\ \emph {et~al.}(2018)\citenamefont {Norcia}, \citenamefont {Lewis-Swan}, \citenamefont {Cline}, \citenamefont {Zhu}, \citenamefont {Rey},\ and\ \citenamefont {Thompson}}]{SpinExchange_Norcia}%
  \BibitemOpen
  \bibfield  {author} {\bibinfo {author} {\bibfnamefont {M.~A.}\ \bibnamefont {Norcia}}, \bibinfo {author} {\bibfnamefont {R.~J.}\ \bibnamefont {Lewis-Swan}}, \bibinfo {author} {\bibfnamefont {J.~R.}\ \bibnamefont {Cline}}, \bibinfo {author} {\bibfnamefont {B.}~\bibnamefont {Zhu}}, \bibinfo {author} {\bibfnamefont {A.~M.}\ \bibnamefont {Rey}},\ and\ \bibinfo {author} {\bibfnamefont {J.~K.}\ \bibnamefont {Thompson}},\ }\href@noop {} {\bibfield  {journal} {\bibinfo  {journal} {Science}\ }\textbf {\bibinfo {volume} {361}},\ \bibinfo {pages} {259} (\bibinfo {year} {2018})}\BibitemShut {NoStop}%
\bibitem [{\citenamefont {Muniz}\ \emph {et~al.}(2020)\citenamefont {Muniz}, \citenamefont {Barberena}, \citenamefont {Lewis-Swan}, \citenamefont {Young}, \citenamefont {Cline}, \citenamefont {Rey},\ and\ \citenamefont {Thompson}}]{PhaseTransition_Muniz}%
  \BibitemOpen
  \bibfield  {author} {\bibinfo {author} {\bibfnamefont {J.~A.}\ \bibnamefont {Muniz}}, \bibinfo {author} {\bibfnamefont {D.}~\bibnamefont {Barberena}}, \bibinfo {author} {\bibfnamefont {R.~J.}\ \bibnamefont {Lewis-Swan}}, \bibinfo {author} {\bibfnamefont {D.~J.}\ \bibnamefont {Young}}, \bibinfo {author} {\bibfnamefont {J.~R.}\ \bibnamefont {Cline}}, \bibinfo {author} {\bibfnamefont {A.~M.}\ \bibnamefont {Rey}},\ and\ \bibinfo {author} {\bibfnamefont {J.~K.}\ \bibnamefont {Thompson}},\ }\href@noop {} {\bibfield  {journal} {\bibinfo  {journal} {Nature}\ }\textbf {\bibinfo {volume} {580}},\ \bibinfo {pages} {602} (\bibinfo {year} {2020})}\BibitemShut {NoStop}%
\bibitem [{\citenamefont {Tavis}\ and\ \citenamefont {Cummings}(1968)}]{TavisCummings}%
  \BibitemOpen
  \bibfield  {author} {\bibinfo {author} {\bibfnamefont {M.}~\bibnamefont {Tavis}}\ and\ \bibinfo {author} {\bibfnamefont {F.~W.}\ \bibnamefont {Cummings}},\ }\href@noop {} {\bibfield  {journal} {\bibinfo  {journal} {Physical Review}\ }\textbf {\bibinfo {volume} {170}},\ \bibinfo {pages} {379} (\bibinfo {year} {1968})}\BibitemShut {NoStop}%
\bibitem [{\citenamefont {Dong}\ \emph {et~al.}(2022)\citenamefont {Dong}, \citenamefont {Zhang}, \citenamefont {Wu},\ and\ \citenamefont {Wu}}]{dong_TavisCummings}%
  \BibitemOpen
  \bibfield  {author} {\bibinfo {author} {\bibfnamefont {Z.}~\bibnamefont {Dong}}, \bibinfo {author} {\bibfnamefont {G.}~\bibnamefont {Zhang}}, \bibinfo {author} {\bibfnamefont {A.-G.}\ \bibnamefont {Wu}},\ and\ \bibinfo {author} {\bibfnamefont {R.-B.}\ \bibnamefont {Wu}},\ }\href@noop {} {\bibfield  {journal} {\bibinfo  {journal} {IEEE Transactions on Automatic Control}\ }\textbf {\bibinfo {volume} {68}},\ \bibinfo {pages} {2048} (\bibinfo {year} {2022})}\BibitemShut {NoStop}%
\bibitem [{\citenamefont {Castin}\ \emph {et~al.}(1989)\citenamefont {Castin}, \citenamefont {Wallis},\ and\ \citenamefont {Dalibard}}]{Doppler_Castin}%
  \BibitemOpen
  \bibfield  {author} {\bibinfo {author} {\bibfnamefont {Y.}~\bibnamefont {Castin}}, \bibinfo {author} {\bibfnamefont {H.}~\bibnamefont {Wallis}},\ and\ \bibinfo {author} {\bibfnamefont {J.}~\bibnamefont {Dalibard}},\ }\href {https://doi.org/10.1364/JOSAB.6.002046} {\bibfield  {journal} {\bibinfo  {journal} {J. Opt. Soc. Am. B}\ }\textbf {\bibinfo {volume} {6}},\ \bibinfo {pages} {2046} (\bibinfo {year} {1989})}\BibitemShut {NoStop}%
\bibitem [{\citenamefont {Bartolotta}(2021)}]{Thesis_Bartolotta}%
  \BibitemOpen
  \bibfield  {author} {\bibinfo {author} {\bibfnamefont {J.}~\bibnamefont {Bartolotta}},\ }\emph {\bibinfo {title} {Advances in Laser Slowing, Cooling, and Trapping using Narrow Linewidth Optical Transitions}},\ \href@noop {} {Ph.D. thesis},\ \bibinfo  {school} {University of Colorado Boulder}, \bibinfo {address} {Boulder} (\bibinfo {year} {2021})\BibitemShut {NoStop}%
\bibitem [{Note1()}]{suppMat}%
  \BibitemOpen
  \bibinfo {note} {See Supplemental Material. Here, we show all the relevant numerical parameters used, eliminate the excited state, second quantize the momentum degree of freedom, eliminate the cavity, put bounds on the two level approximation, and optimize the spin squeezing parameter through semi-classical treatment.}\BibitemShut {Stop}%
\bibitem [{\citenamefont {J{\"a}ger}\ \emph {et~al.}(2022)\citenamefont {J{\"a}ger}, \citenamefont {Schmit}, \citenamefont {Morigi}, \citenamefont {Holland},\ and\ \citenamefont {Betzholz}}]{Lindblad_jager}%
  \BibitemOpen
  \bibfield  {author} {\bibinfo {author} {\bibfnamefont {S.~B.}\ \bibnamefont {J{\"a}ger}}, \bibinfo {author} {\bibfnamefont {T.}~\bibnamefont {Schmit}}, \bibinfo {author} {\bibfnamefont {G.}~\bibnamefont {Morigi}}, \bibinfo {author} {\bibfnamefont {M.~J.}\ \bibnamefont {Holland}},\ and\ \bibinfo {author} {\bibfnamefont {R.}~\bibnamefont {Betzholz}},\ }\href@noop {} {\bibfield  {journal} {\bibinfo  {journal} {Physical Review Letters}\ }\textbf {\bibinfo {volume} {129}},\ \bibinfo {pages} {063601} (\bibinfo {year} {2022})}\BibitemShut {NoStop}%
\bibitem [{\citenamefont {Yukawa}\ and\ \citenamefont {Nemoto}(2016)}]{SUN_Yukawa}%
  \BibitemOpen
  \bibfield  {author} {\bibinfo {author} {\bibfnamefont {E.}~\bibnamefont {Yukawa}}\ and\ \bibinfo {author} {\bibfnamefont {K.}~\bibnamefont {Nemoto}},\ }\href {https://doi.org/10.1088/1751-8113/49/25/255301} {\bibfield  {journal} {\bibinfo  {journal} {Journal of Physics A: Mathematical and Theoretical}\ }\textbf {\bibinfo {volume} {49}},\ \bibinfo {pages} {255301} (\bibinfo {year} {2016})}\BibitemShut {NoStop}%
\bibitem [{\citenamefont {Davis}\ \emph {et~al.}(2020)\citenamefont {Davis}, \citenamefont {Periwal}, \citenamefont {Cooper}, \citenamefont {Bentsen}, \citenamefont {Evered}, \citenamefont {Van~Kirk},\ and\ \citenamefont {Schleier-Smith}}]{GapProtection_SchleierSmith}%
  \BibitemOpen
  \bibfield  {author} {\bibinfo {author} {\bibfnamefont {E.~J.}\ \bibnamefont {Davis}}, \bibinfo {author} {\bibfnamefont {A.}~\bibnamefont {Periwal}}, \bibinfo {author} {\bibfnamefont {E.~S.}\ \bibnamefont {Cooper}}, \bibinfo {author} {\bibfnamefont {G.}~\bibnamefont {Bentsen}}, \bibinfo {author} {\bibfnamefont {S.~J.}\ \bibnamefont {Evered}}, \bibinfo {author} {\bibfnamefont {K.}~\bibnamefont {Van~Kirk}},\ and\ \bibinfo {author} {\bibfnamefont {M.~H.}\ \bibnamefont {Schleier-Smith}},\ }\href {https://doi.org/10.1103/PhysRevLett.125.060402} {\bibfield  {journal} {\bibinfo  {journal} {Physical Review Letters}\ }\textbf {\bibinfo {volume} {125}},\ \bibinfo {pages} {060402} (\bibinfo {year} {2020})}\BibitemShut {NoStop}%
\bibitem [{\citenamefont {Hu}\ \emph {et~al.}(2017)\citenamefont {Hu}, \citenamefont {Chen}, \citenamefont {Vendeiro}, \citenamefont {Urvoy}, \citenamefont {Braverman},\ and\ \citenamefont {Vuleti\ifmmode~\acute{c}\else \'{c}\fi{}}}]{VacuumSqueezing}%
  \BibitemOpen
  \bibfield  {author} {\bibinfo {author} {\bibfnamefont {J.}~\bibnamefont {Hu}}, \bibinfo {author} {\bibfnamefont {W.}~\bibnamefont {Chen}}, \bibinfo {author} {\bibfnamefont {Z.}~\bibnamefont {Vendeiro}}, \bibinfo {author} {\bibfnamefont {A.}~\bibnamefont {Urvoy}}, \bibinfo {author} {\bibfnamefont {B.}~\bibnamefont {Braverman}},\ and\ \bibinfo {author} {\bibfnamefont {V.}~\bibnamefont {Vuleti\ifmmode~\acute{c}\else \'{c}\fi{}}},\ }\href {https://doi.org/10.1103/PhysRevA.96.050301} {\bibfield  {journal} {\bibinfo  {journal} {Phys. Rev. A}\ }\textbf {\bibinfo {volume} {96}},\ \bibinfo {pages} {050301} (\bibinfo {year} {2017})}\BibitemShut {NoStop}%
\bibitem [{\citenamefont {Reilly}\ \emph {et~al.}(2024)\citenamefont {Reilly}, \citenamefont {J{\"a}ger}, \citenamefont {Wilson}, \citenamefont {Cooper}, \citenamefont {Eggert},\ and\ \citenamefont {Holland}}]{PDD_Reilly}%
  \BibitemOpen
  \bibfield  {author} {\bibinfo {author} {\bibfnamefont {J.~T.}\ \bibnamefont {Reilly}}, \bibinfo {author} {\bibfnamefont {S.~B.}\ \bibnamefont {J{\"a}ger}}, \bibinfo {author} {\bibfnamefont {J.~D.}\ \bibnamefont {Wilson}}, \bibinfo {author} {\bibfnamefont {J.}~\bibnamefont {Cooper}}, \bibinfo {author} {\bibfnamefont {S.}~\bibnamefont {Eggert}},\ and\ \bibinfo {author} {\bibfnamefont {M.~J.}\ \bibnamefont {Holland}},\ }\href@noop {} {\bibfield  {journal} {\bibinfo  {journal} {Physical Review Research}\ }\textbf {\bibinfo {volume} {6}},\ \bibinfo {pages} {033090} (\bibinfo {year} {2024})}\BibitemShut {NoStop}%
\bibitem [{\citenamefont {Luo}\ \emph {et~al.}(2024{\natexlab{b}})\citenamefont {Luo}, \citenamefont {Zhang}, \citenamefont {Chu}, \citenamefont {Maruko}, \citenamefont {Rey},\ and\ \citenamefont {Thompson}}]{TACT_luo}%
  \BibitemOpen
  \bibfield  {author} {\bibinfo {author} {\bibfnamefont {C.}~\bibnamefont {Luo}}, \bibinfo {author} {\bibfnamefont {H.}~\bibnamefont {Zhang}}, \bibinfo {author} {\bibfnamefont {A.}~\bibnamefont {Chu}}, \bibinfo {author} {\bibfnamefont {C.}~\bibnamefont {Maruko}}, \bibinfo {author} {\bibfnamefont {A.~M.}\ \bibnamefont {Rey}},\ and\ \bibinfo {author} {\bibfnamefont {J.~K.}\ \bibnamefont {Thompson}},\ }\href@noop {} {\bibfield  {journal} {\bibinfo  {journal} {arXiv preprint arXiv:2402.19429}\ } (\bibinfo {year} {2024}{\natexlab{b}})}\BibitemShut {NoStop}%
\bibitem [{\citenamefont {Shankar}\ \emph {et~al.}(2019{\natexlab{b}})\citenamefont {Shankar}, \citenamefont {Greve}, \citenamefont {Wu}, \citenamefont {Thompson},\ and\ \citenamefont {Holland}}]{ContinuousPhase_Shankar}%
  \BibitemOpen
  \bibfield  {author} {\bibinfo {author} {\bibfnamefont {A.}~\bibnamefont {Shankar}}, \bibinfo {author} {\bibfnamefont {G.~P.}\ \bibnamefont {Greve}}, \bibinfo {author} {\bibfnamefont {B.}~\bibnamefont {Wu}}, \bibinfo {author} {\bibfnamefont {J.~K.}\ \bibnamefont {Thompson}},\ and\ \bibinfo {author} {\bibfnamefont {M.}~\bibnamefont {Holland}},\ }\href {https://doi.org/10.1103/PhysRevLett.122.233602} {\bibfield  {journal} {\bibinfo  {journal} {Phys. Rev. Lett.}\ }\textbf {\bibinfo {volume} {122}},\ \bibinfo {pages} {233602} (\bibinfo {year} {2019}{\natexlab{b}})}\BibitemShut {NoStop}%
\bibitem [{\citenamefont {Steck}(2001)}]{Rb_Steck}%
  \BibitemOpen
  \bibfield  {author} {\bibinfo {author} {\bibfnamefont {D.~A.}\ \bibnamefont {Steck}},\ }\href {https://steck.us/alkalidata/rubidium87numbers.pdf} {\bibinfo {title} {Rubidium 87 d line data}} (\bibinfo {year} {2001})\BibitemShut {NoStop}%
\bibitem [{\citenamefont {Steck}(2007)}]{QO_Steck}%
  \BibitemOpen
  \bibfield  {author} {\bibinfo {author} {\bibfnamefont {D.~A.}\ \bibnamefont {Steck}},\ }\href {http://steck.us/teaching} {\bibinfo {title} {Quantum and atom optics}} (\bibinfo {year} {2007})\BibitemShut {NoStop}%
\bibitem [{\citenamefont {Lewis-Swan}\ \emph {et~al.}(2018)\citenamefont {Lewis-Swan}, \citenamefont {Norcia}, \citenamefont {Cline}, \citenamefont {Thompson},\ and\ \citenamefont {Rey}}]{RobustSpinSqueeze_Rey}%
  \BibitemOpen
  \bibfield  {author} {\bibinfo {author} {\bibfnamefont {R.~J.}\ \bibnamefont {Lewis-Swan}}, \bibinfo {author} {\bibfnamefont {M.~A.}\ \bibnamefont {Norcia}}, \bibinfo {author} {\bibfnamefont {J.~R.~K.}\ \bibnamefont {Cline}}, \bibinfo {author} {\bibfnamefont {J.~K.}\ \bibnamefont {Thompson}},\ and\ \bibinfo {author} {\bibfnamefont {A.~M.}\ \bibnamefont {Rey}},\ }\href {https://doi.org/10.1103/PhysRevLett.121.070403} {\bibfield  {journal} {\bibinfo  {journal} {Phys. Rev. Lett.}\ }\textbf {\bibinfo {volume} {121}},\ \bibinfo {pages} {070403} (\bibinfo {year} {2018})}\BibitemShut {NoStop}%
\end{thebibliography}
\end{document}